 \documentclass[final,3p,times,twocolumn]{elsarticle}

\usepackage{amssymb}
\usepackage{amsmath}

\usepackage{lineno}
\usepackage{xurl,hyperref}
\usepackage{tikz}
\usepackage{float}
\usetikzlibrary{mindmap,trees,shadows} 
\usepackage[most]{tcolorbox}

\newtcbtheorem{Summary}{\bfseries Summary}{enhanced,drop shadow={black!50!white},
  coltitle=black,
  top=0.3in,
  attach boxed title to top left=
  {xshift=1.5em,yshift=-\tcboxedtitleheight/2},
  boxed title style={size=small,colback=pink}
}{summary}

\newtcolorbox[auto counter]{Insight}[1][]{title={\bfseries Insight~\thetcbcounter},enhanced,drop shadow={black!50!white},
  coltitle=black,
  top=0.1in,
  attach boxed title to top left=
  {xshift=1.5em,yshift=-\tcboxedtitleheight/2},
  boxed title style={size=small,colback=pink},#1}

\journal{Nuclear Physics B}

\begin{document}

\begin{frontmatter}



\title{Brain Imaging Foundation Models, Are We There Yet? A Systematic Review of Foundation Models for Brain Imaging and Biomedical Research }


\author[lih]{Salah GHAMIZI\corref{cor1}}
\ead{salah.ghamizi@lih.lu}

\author[lih]{Georgia KANLI}
\ead{georgia.kanli@lih.lu}

\author[kcl]{Yu DENG}
\ead{yu.deng@kcl.ac.uk}

\author[lih]{Magali PERQUIN}
\ead{magali.perquin@lih.lu}

\author[lih]{Olivier KEUNEN}
\ead{olivier.keunen@lih.lu}

\address[lih]{Luxembourg Institute of Health (LIH)}
\address[kcl]{King's College London, UK}

\cortext[cor1]{Corresponding author}

\begin{abstract}
Foundation models (FMs), large neural networks pretrained on extensive and diverse datasets, have revolutionized artificial intelligence and shown significant promise in medical imaging by enabling robust performance with limited labeled data. Although numerous surveys have reviewed the application of FM in healthcare care, brain imaging remains underrepresented, despite its critical role in the diagnosis and treatment of neurological diseases using modalities such as MRI, CT, and PET. Existing reviews either marginalize brain imaging or lack depth on the unique challenges and requirements of FM in this domain, such as multimodal data integration, support for diverse clinical tasks, and handling of heterogeneous, fragmented datasets.

To address this gap, we present the first comprehensive and curated review of FMs for brain imaging. We systematically analyze 161 brain imaging datasets and 86 FM architectures, providing information on key design choices, training paradigms, and optimizations driving recent advances. Our review highlights the leading models for various brain imaging tasks, summarizes their innovations, and critically examines current limitations and blind spots in the literature. We conclude by outlining future research directions to advance FM applications in brain imaging, with the aim of fostering progress in both clinical and research settings.
\end{abstract}



\begin{keyword}
brain imaging \sep deep learning \sep foundation models \sep brain cancer \sep neurodegenerative diseases \sep neurovascular diseases 


\end{keyword}

\end{frontmatter}



\section{Introduction}
\label{sec-intro}
In recent years, foundation models (FMs), large neural networks pre-trained on very large and diverse datasets, have greatly advanced the field of artificial intelligence (AI) \cite{Kolides2023, Moor2023, Schneider2024, Mahesh2024,Huang2025, Bian2025}.
They offer several benefits in medical imaging, including adaptability with fewer labeled examples, modularity, and robustness. In medical imaging, FMs can significantly reduce the demand for training samples, which is particularly advantageous in scenarios where large labeled datasets are scarce\cite{eslami_pubmedclip_2023,cheng_sam-med2d_2023}. 

With the increasing popularity of FMs, dozens of surveys have covered their application in healthcare \cite{moor2023foundation,azad_foundational_2023,qiu2023large,thirunavukarasu2023large,he_foundation_2024,shrestha2023medical,wang2023pretrained,zhou2023survey,yuan2023large,lee2024foundation,li2024progress,liu2024large,qiu2023pretraining,wang2023accelerating,zhang2024challenges,zhang2024datacentric,zhang_segment_2024,zhao_clip_2024,shi_survey_2024,he2023survey}. However, none of them focused on brain FMs, leading to a large disregard for the models tackling this critical field. For example, \citep{he_foundation_2024}, one of the largest reviews included 200 technical papers, but only 32 covered brain FMs; Similarly, the study of \citet{shi_survey_2024} provided one of the most thorough analysis and covered 76 research articles, but only ten dealt with brain imaging. Even surveys on a specific family of FMs largely ignored FMs for brain imaging. The study of \citet{zhao_clip_2024}, the largest review of CLIP-based FMs, only covered seven brain FMs, while \citet{zhang_segment_2024} who explored SAM-based FMs identified only nine brain FMs. 

In summary, despite the growing body of research on FMs for medical imaging, existing surveys often either underrepresent brain imaging by focusing on it as just one of many organs studied or fail to provide in-depth insight into the specificities of FMs for brain pathologies. 
First, brain pathologies often require the analysis of various data sources, including MRI, CT, PET scans, genetic information, and clinical histories, with some time series (fMRI) or longitudinal data. Next, FM models should support a large set of tasks in a brain setting, from segmentation to classification, contrast synthesis, and even question answering. Brain research also poses multiple pathology-specific challenges. Each type of pathology involves a different area of the brain, requires different imaging protocols, expertise, and analysis. Finally, brain imaging datasets are scattered and require specific legal and technical processing protocols that hinder their distribution, experiment replication, and benchmarking.

In fact, brain imaging has become very important in modern medicine. 
Techniques such as Magnetic Resonance Imaging (MRI), Computed Tomography (CT), and Positron Emission Tomography (PET) have improved our understanding of brain problems such as tumors, strokes, Alzheimer's disease, Parkinson's disease, and multiple sclerosis \cite{aljahdali2024effectiveness, Kanli2025, Keunen2011, Sauvage2022, Kontopodis2015, Zhou2022}. Continuous advancements in imaging technologies also aim to improve image quality, providing clearer and more accurate views of these diseases \cite{Boudissa2024, Perlo2025}. MRI is especially valuable because it produces detailed images of brain structures. It is widely used in both research and medical care.

Despite the importance of brain imaging, to the best of our knowledge, no review has focused on FMs for brain imaging applications. To fill this gap, we provide a focused review of FM in brain image analysis. Our goal is to provide a clear picture of current progress, trends, and challenges. We organize this review into four parts:
(1) An introduction to critical concepts and definitions related to FM brain imaging (Section \ref{sec-definitions}). 
(2) A summary of the methodology of our review (Section \ref{sec-methodology}). 
(3) A synthesis of the main insights from our extensive review of brain imaging FMs (Section \ref{sec-findings}), from the review of brain imaging datasets (Section \ref{sec-datasets}), and which FMs are most relevant per task and dataset (Section \ref{sec:best-archs}).
(4) A discussion of the main directions and trends in brain FM research and the main pitfalls and blind spots of the current literature (Section \ref{sec-limitations}).

In summary, our contributions are as follows.

\begin{itemize}
\item We propose the first highly curated review of FM for brain imaging, in which we collected 161 brain imaging datasets and analyzed 86 models.

\item We identify the main design choices, training paradigms, and optimizations that led to recent leaps in brain FM performance and efficiency.

\item We highlight some of the best FM for each use case and summarize their main innovations and contributions.

\item We discuss the main limitations of FM research for brain imaging and suggest the main directions to investigate.
\end{itemize}

\section{Concepts \& Definitions}
\label{sec-definitions}

Foundation models (FM)s are large-scale machine learning models trained on extensive datasets to perform a wide range of tasks through adaptation. They serve as versatile starting points for developing specialized AI applications, enabling efficiency and scalability compared to traditional task-specific models. The term was coined by~\citet{bommasani2021opportunities} and refers to models that meet the following criteria:

\begin{itemize}
    \item \textbf{Broad training data}: The models are trained on massive and diverse datasets (often unlabeled and unstructured) to capture general patterns.	
    \item \textbf{Self-supervised learning}: The training uses self-supervised or unsupervised techniques.	
    \item \textbf{Adaptability}: The evaluation demonstrates the effectiveness of specialized downstream tasks with fine-tuning.
\end{itemize}

Regulations also provide requirements. For example, the 2023 US AI FM Transparency Act\footnote{\url{https://beyer.house.gov/uploadedfiles/one-pager_ai_foundation_model_transparency_act_.pdf}} requires that FM contain at least one billion parameters. 

We build on the requirements proposed by ~\citet{bommasani2021opportunities} and further clarify what is meant by \emph{diverse datasets} \emph{downstream tasks} in our setting, thus defining the scope of our review. We restricted our study to FMs that support medical imaging data, among other modalities.

\subsection{Downstream tasks} Brain FM are generally effective in more than one task with zero-shot learning (generalization without retraining), or with light fine-tuning. The tasks can be segmentation tasks (marked \textbf{S}), generation tasks (marked \textbf{G}), classification tasks (marked \textbf{C}), or regression tasks (marked \textbf{R}). We consider the following tasks in our study:

\begin{enumerate}
    \item \textbf{Structure parcellation (S)}: Divides the brain into functionally or structurally distinct regions based on properties like connectivity, activity patterns, or microstructural features.
    \item \textbf{Tissue segmentation (S)}: Outputs a segmentation of brain voxels into anatomical tissue types—gray matter (GM), white matter (WM), and cerebrospinal fluid (CSF).
    \item \textbf{Anomaly segmentation (S)}: Outputs a segmentation of brain voxels into pathological regions (e.g., tumors, lesions).
    \item \textbf{Anomaly classification (C)}: Categorizes detected abnormalities (e.g., glioma subtypes, stroke types) based on imaging features.
    \item \textbf{Modality classification (C)}: Recognizes MRI sequence types (e.g. T1w, FLAIR) from intensity patterns and anatomical contrasts.
    \item \textbf{Survival prediction (R)}: Estimates patient prognosis (e.g., glioma survival timelines or expected lifespan) using imaging biomarkers.
    \item \textbf{Age prediction (R)}: Predicts chronological age from brain structure; with deviations between predicted age and actual age possibly indicating a risk of neurodegeneration.
    \item \textbf{Report generation (G)}: Automates radiology report drafting by synthesizing findings following a pre-defined template.
    \item \textbf{Question answering (G)}: Answers clinical queries about scans using multimodal image-text understanding.
    \item \textbf{Image retrieval (G)}: Retrieves similar historical cases (images or reports) by matching multimodal image-text features.
    \item \textbf{Image synthesis (G)}: Generates synthetic scans (e.g. pseudo-healthy reconstructions) or sequence translations (e.g. T1W $\rightarrow$ T2W).
    \item \textbf{Image registration (G)}: Aligns multi-modal or longitudinal scans to a common anatomical space.
    \item \textbf{Super-resolution (G)}: Enhances the resolution of low-quality scans to higher resolutions.
    \item \textbf{Motion removal (G)}: Corrects motion artifacts in MRI by disentangling anatomical content from movement.
    \item \textbf{Image denoising (G)}: Reduces noise in acquisitions with low-Signal-to-Noise Ratio (SNR) while preserving structural details.
    \item \textbf{Contrast improvement (G)}: Optimizes tissue visibility by adjusting intensity distributions across sequences.
    \item \textbf{High field translation (G)}: Simulates high-field MRI characteristics from low-field inputs (e.g., 7 Tesla from 1,5 or 3 Tesla) .
    \item \textbf{Harmonization (G)}: Standardizes scan appearance across scanners/sites/protocols for multi-center studies comparability.
\end{enumerate}

\subsection{Medical imaging jargon} Terminology used in the medical and AI engineering fields may sometimes differ and create confusion. To clarify reading, we have followed in this text the jargon of the Cancer Imaging Archive (TCIA)\footnote{\url{https://www.cancerimagingarchive.net/}}.
Brain imaging FMs can be split into two categories: Image-Language FMs, which accept both text and image inputs (referred to as multi-modal models), and Image-only FMs, which process only images (referred to as single-modal models). 

\paragraph{Imaging modalities}

The most popular brain imaging modalities are Magnetic Resonance Imaging (MRI) and Computer Tomography (CT). 

MRI uses strong magnetic fields and radio waves to generate high-resolution images of soft tissues. Multiple MRI sequences are available: T1-Weighted (T1w) sequences highlight fat and anatomical details and are useful for structural assessment. T2-Weighted (T2w) sequences emphasize fluids and pathological tissue and are commonly used to detect edema or lesions. Fluid-Attenuated Inversion Recovery (FLAIR) sequences suppress the signal from cerebrospinal fluid (CSF) to better visualize lesions near ventricles. Diffusion-Weighted Imaging (DWI) and Diffusion Tensor Imaging (DTI) sequences are sensitive to water motion. They are used to detect acute ischemic stroke, assess cellular density and reconstruct fiber tracts. Susceptibility Weighted Imaging (SWI) detects blood products and microbleeds, while Magnetic Resonance Angiography (MRA) makes it possible to visualize blood vessels. Gradient Echo T2*-Weighted sequences (T2*) are used for detecting hemorrhages and microbleeds, and functional MRI (fMRI) maps brain activity by detecting changes in blood oxygenation during tasks or at rest. T1 and T2 sequences can make use of contrast agents. For example, T1-weighted Contrast-Enhanced Imaging (T1ce) images are acquired after intravenous injection of a Gadolinium-based contrast agent, enhancing visualization of blood vessels, tumors, and areas with abnormal blood-brain barrier.

CT uses X-ray imaging to produce cross-sectional views of the brain. It is commonly used to detect bleeding, skull fractures, and swelling. Computer Tomography Angiography (CTA) repeatedly acquire CT scans during and after intravenous  injection of a contrast agent, to provide blood perfusion maps of the brain and neck, helping diagnose vascular conditions like aneurysms, stenosis, or dissections.

Positron Emission Tomography (PET) involves injection of a positron-emitting radioactive tracer to measure metabolic activity and various molecular biomarkers in brain tissue, useful for diagnosing brain pathologies. Single Photon Emission CT (SPECT) provides similar functions using gamma ray-emitting radioisotopes. These nuclear medicine techniques provide access to various physiological and molecular readouts based on the tracer inserted in the body prior to the scan. For instance, PET uses \textsuperscript{18}F-labelled tracers such as FDG to assess brain metabolism or AV45 to assess amyloid beta plaques accumulation in Alzheimer Disease (AD), and SPECT uses \textsuperscript{123}Iodine or \textsuperscript{99m}Technetium labelled tracers to assess dopamine transport in Parkinson Disease (PD).

Overall, we define as \emph{imaging modality} a specific technology or method (e.g. MRI, CT, PET, SPECT, Ecograph) that generates diagnostic images. 
Each imaging modality can make use of different types of \emph{sequences} or acquisition parameters, like T1w, T2w, ...

\paragraph{Studies, series and sequences}

A \emph{study} represents a single imaging session or examination for a patient, possibly containing multiple series. A \emph{series} is a set of acquisitions within a study corresponding to one sequence or set of parameters. 
The \emph{slices} are the individual pictures that make up a series.

The dataset resulting from a series acquisition can be in 2D, in which case each input is termed a slice or section. Datasets can alternatively be in 3D with inputs termed as volumes, or in 4D with inputs corresponding to volumes acquired over time or another contrast modifying dimension.

A \emph{single-sequence} FM refers to a model that was trained to accept only one type of sequence as input (f.i. T1w), while a \emph{multi-sequence} FM supports multiple sequences. It can accept one type of sequence at a time using 3D tensors as inputs, or multiple sequences combined in one series using 4D inputs. These 4D models sometimes support missing sequences, and their implementation then requires handling 4D inputs with different dimensions per input.   


\subsection{Dataset diversity}
\label{sec:datasets}
We consider a brain medical imaging model to be a FM if it meets at least one of the following criteria: 

-The FM is trained on datasets that include at least two imaging modalities (such as MRI, CT, PET, etc);

-The FM is trained on datasets with medical images from at least two different organs, including the brain;

-The FM is trained on datasets with at least two different brain pathologies (e.g. brain tumor, vascular disease, etc).

For example, models that are trained only on brain images but support MRI and CT inputs for brain tumor diagnosis are considered FMs. Models that used brain and torso tumor datasets are also considered FMs. Similarly, we consider models trained to segment both brain tumors and MS lesions as FMs. 

In the following, we describe the main brain pathologies by prevalence and present the imaging protocols that are commonly used to diagnose it. 

\paragraph{Headache disorders} such as \emph{Tension-Type Headache (TTH)} and \emph{migraines} are the most prevalent neurological disorders globally \cite{Headache2024, GBD2021}. These disorders rely on clinical diagnoses based on history and physical examination rather than neuroimaging findings.

\paragraph{Vascular diseases} These include pathologies such as \emph{strokes} that are the leading cause of adjusted life years for neurological disability (DALY) (42.2\% of total) \cite{StrokeReport2025}. \emph{Ischemic stroke} constitutes 65.3\% of incidents and refers to reduced blood flow to brain tissue. It occurs when a blood clot blocks an artery, cutting off oxygen supply and leading to brain cell damage or death. Ischemic strokes are generally confirmed by DWI MRI. \emph{Hemorrhagic strokes} are less frequent and diagnosed with CT. Other vascular diseases include \emph{cerebral aneurysm} and \emph{artery stenosis}. Cerebral aneurysms are weakness in the walls of brain arteries that causes dilation that is diagnosed by CTA for the detection of rupture and by MRI / MRA for non-ruptured cases. Artery stenosis is a narrowing of the brain arteries (e.g. carotid) due to atherosclerotic plaque that is diagnosed by CTA or MRA.

\paragraph{Neurodegenerative Diseases}
\emph{Parkinson disease (PD)} is a dopaminergic neuron degeneration which prevalence reached 11.77 million cases in 2021\cite{Parkinson2025}. MRI rules out mimics and PET can be used to evaluate dopamine loss. 
\emph{Alzheimer's Disease (AD)} and other dementias affect 56.9 million people globally in 2021\cite{Alzheimers2025}. Alzheimer’s disease manifests through the accumulation of abnormally folded proteins into extracellular amyloid plaques and intracellular neurofibrillary tangles.
MRI can identify brain structures atrophy that are associated with AD, but only PET differentiates protein accumulations. 

\paragraph{Brain tumors} \emph{Metastases} are the most common brain tumors\cite{pmc9533228}, with an estimated incidence of 24.2 per 100,000 persons annually. Anatomical MRI is generally sufficient for detecting metastatic lesions.
\emph{Meningioma} is the most common primary non-malignant brain tumor\cite{cbtrus2023}. CE-MRI is the gold standard for its diagnosis. DOTATATE-PET helps differentiate meningiomas from other dural-based lesions. 
\emph{Glioblastoma Multiforme (GBM)} is the most aggressive primary malignant brain tumor\cite{cbtrus2023}. T1w, T2w, FLAIR, diffusion and perfusion imaging can collectively be used to assess necrosis, cellularity, and blood-brain barrier disruption, while amino acid PET is needed to distinguish tumor recurrence from radiation necrosis.
Other notable brain cancers include \emph{Diffuse Midline Glioma (DMG)}, \emph{Low-Grade Glioma (LGG)}, \emph{Pituitary Adenoma}, \emph{Vestibular Schwannoma} and \emph{Lymphoblastic Lymphoma}. Anatomical MRI is generally sufficient for diagnosis.

\paragraph{Psychiatric disorders} Mood disorders such as \emph{depression} are studied with MRI to link structural changes (e.g., prefrontal cortex) to the clinical diagnosis. \emph{Bipolar disorders} are also characterized by extreme mood swings. Similarly, MRI is used to identify prefrontal/limbic abnormalities that can be associated with the clinical diagnosis. \emph{ADHD, or Attention-Deficit/Hyperactivity Disorder} is also clinically diagnosed, but MRI is used to explore the basal ganglia / prefrontal differences that are linked to this disorder.
\emph{Schizophrenia} is a psychotic disorder in which MRI research shows volume changes, while PET studies the dopamine dysfunction that characterizes such disorders.

\paragraph{Other Neurological Disorders} A major family of neurological disorders is d\emph{emyelinating diseases}. They refer to damage to myelin (e.g., Multiple Sclerosis MS). T2 and FLAIR sequences are used to show white matter lesions, while PET can be used to detect inflammation. \emph{MS lesions} in particular, represent focal demyelination in the brain. T2w sequences identify lesions as white spots and PET (e.g. TSPO tracers) is also used to track microglial activity. Other white matter diseases can be related to aging and are primary diagnosed with MRI (T2/FLAIR hyperintensities).  
\emph{Attention Deficit Hyperactivity Disorder (ADHD)} is the most common neurodevelopmental disorder in children and adolescents
ADHD affects 84.8 million people worldwide \cite{ADHD2025}. ADHD cannot be diagnosed solely with medical imaging, even if altered activation patterns on functional MRI (fMRI), and changes in dopamine metabolism on PET scans can reveal patterns associated with ADHD.
Another common disorder is \emph{epilepsy}. It affects 51.7 million people globally \cite{Epilepsy2025} and is characterized by recurrent seizures due to abnormal electrical activity in the brain. Medical imaging can be used for diagnosis, with MRI used to detect lesions (e.g. hippocampal sclerosis), and PET / SPECT to localize the focus of the seizure. Another prevalent spectrum disorder is \emph{autism} (61.8 million people globally \cite{Autism2025}). MRI can help study structural and functional connectivity, but is not used for diagnosis.




In general, brain imaging datasets (and thus brain models) can be used to diagnose and study a wide set of pathologies and disorders. 
Although diverse, these disorders share common diagnostic protocols that motivate the development and distribution of general FMs dedicated to brain research.

\section{Methodology}
\label{sec-methodology}
\begin{figure*}[t]
    \centering
    \includegraphics[trim=0 11em 0 2em, clip, width=\linewidth]{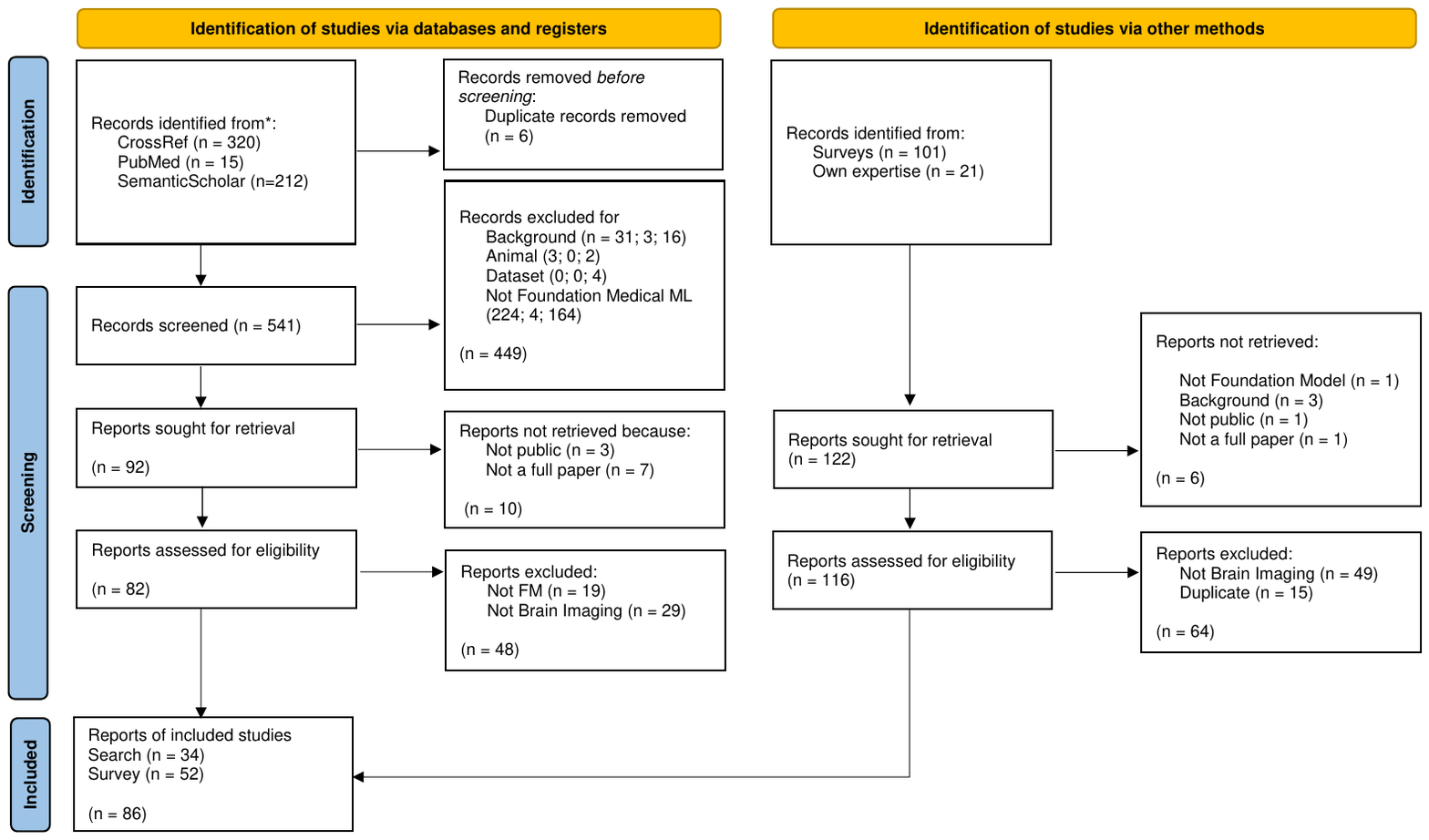}
    \caption{Prisma Methodology followed in our SLR. The databases and registries are Scholar, Crossref, and PubMed; The other methods include papers identified through previous surveys and snowballing from publications found in the registries.}
    \label{fig:prisma-methodology}
\end{figure*}

Our review follows the Systematic Literature Review (SLR) methodology as introduced by Kitchenham~\cite{kitchenham2004procedures}, ensuring a structured and reproducible approach to identifying and analyzing foundation deep learning models for brain imaging. We adhere to the PRISMA 2020 guidelines~\cite{page2021prisma} throughout the selection and evaluation process.

We present in Fig. \ref{fig:prisma-methodology} the steps of our data collection and screening metholodgy. Our publication collection is two-fold: We first collected relevant publications using the standardized search strategy below, then we enriched the publication list through recent surveys and snowballing. 

\subsection{Search strategy}

We built the publications listing on February 18, 2025 using the respective Application Programming Interface (API) of three research search engine: SemanticScholar, Crossref, and PubMed. These providers offer a free API with powerful querying abilities, including abstract search with exact expressions. While Semantic Scholar and Crossref provide an exhaustive view, PubMed focuses primarily on the biomedical and life sciences literature. These search engines  not only cover major journal and conference peer-reviewed publications, but also support preprints from arXiv and medRxiv for e.g.

Each API offers its unique way of querying, with \emph{and} and \emph{or} operators handled separately. 

We adapted the query to each search engine. We can express the query in plain language as searching for all publications that contain in their abstract or title keywords related to FMs for brain. The keywods should include either "FM" or "large model" or "large vision model". We exclude publications that mention in their abstract or title reviews, surveys and animal research keywords. In particular, we explicitly excluded pig and mice keywords as they represent the most common animal studies that the queries were retrieving. The abstract or title should include "brain" and one of thee medical imaging expressions "MR" or "MRI" or "CT" or "PET" or "medical image" or "medical imaging".

Our query formatted for Semantic Scholar is as follows: \texttt{'("FM" | "large model" | "large vision model") brain -animal -mice -survey -pig -review (MR|MRI|CT|PET|"medical image"|"medical imaging")'}. 

We restrict our research to the last 5 years. This initial search yielded 212, 320, and 15 results on Semantic Scholar, Crossref and PubMed respectively (left pipeline in Fig. \ref{fig:prisma-methodology}). 

By manually checking the titles and abstracts, we excluded 449 publications, and attempted to collect the full papers among the 92 remaining. The excluded publications were mostly (392) because the papers did not introduce new deep learning models. Among these 392 excluded publications, 224 were found through CrossRef, 164 through Semantic Scholars, and four through PubMed. 

When retrieving the 92 remaining, three were not accessible and the authors did not provide us with full papers in time, and seven were actually posters or abstract papers. The remaining 82 papers underwent a full-text assessment to ensure they met the predefined inclusion criteria. To be included in our SLR a publication needed to (1) introduce a novel brain FM, following our earlier definition (section \ref{sec-definitions}), and (2) have been evaluated on at least one dataset with brain/head imaging area of interest. 19 publications did not meet the first criteria and 29 did not meet the second. In conclusion, 34 research papers were included in this systematic review analysis.

\subsection{Augmentation strategy}

We extended our database with citations from two additional sources. We included 101 publications from the last five peer-reviewed surveys/reviews of FMs for medical applications ~\cite{shi_survey_2024,he_foundation_2024,azad_foundational_2023,zhao_clip_2024,zhang_segment_2024}. We then added 21 publications that were previously cited in the collected papers (snowballing). The surveys and snowballing cited more publications, but we included only the ones where the title hinted to publications that fall under the scope of our study.

From title and abstract reading, we did not collect six publications among the 122 for various reasons (right pipeline in Fig. \ref{fig:prisma-methodology}): one publication introduced a new model, that did not match our definition of FM, three were about FMs, but did not introduce new models, and focused on comparing or studying existing models. One publication was not available for retrieval, and one publication was only an abstract and poster.  

After a full-text assessment following the same inclusion criteria as the search step, we obtain 52 new research papers that we include in this systematic review analysis.  

\subsection{Analysis synthesis}

We analyzed the publications both using their meta-data (authors, venues, publication years), and following the research questions of our study. 
We share publicly the detailed assessment in an interative tabular visualization on Notion \url{https://ballistic-purple-6f5.notion.site/182bf710ca5980d6b3a4fd3402960f0e?v=fd73fbce802146559cefa7fc0f679d26}.  

\section{Findings: Brain Foundation Models}
\label{sec-findings}

\subsection{General trends}

\begin{figure}[t]
    \centering
    \includegraphics[trim=0em 0em 0em 0em, clip, width=\linewidth]{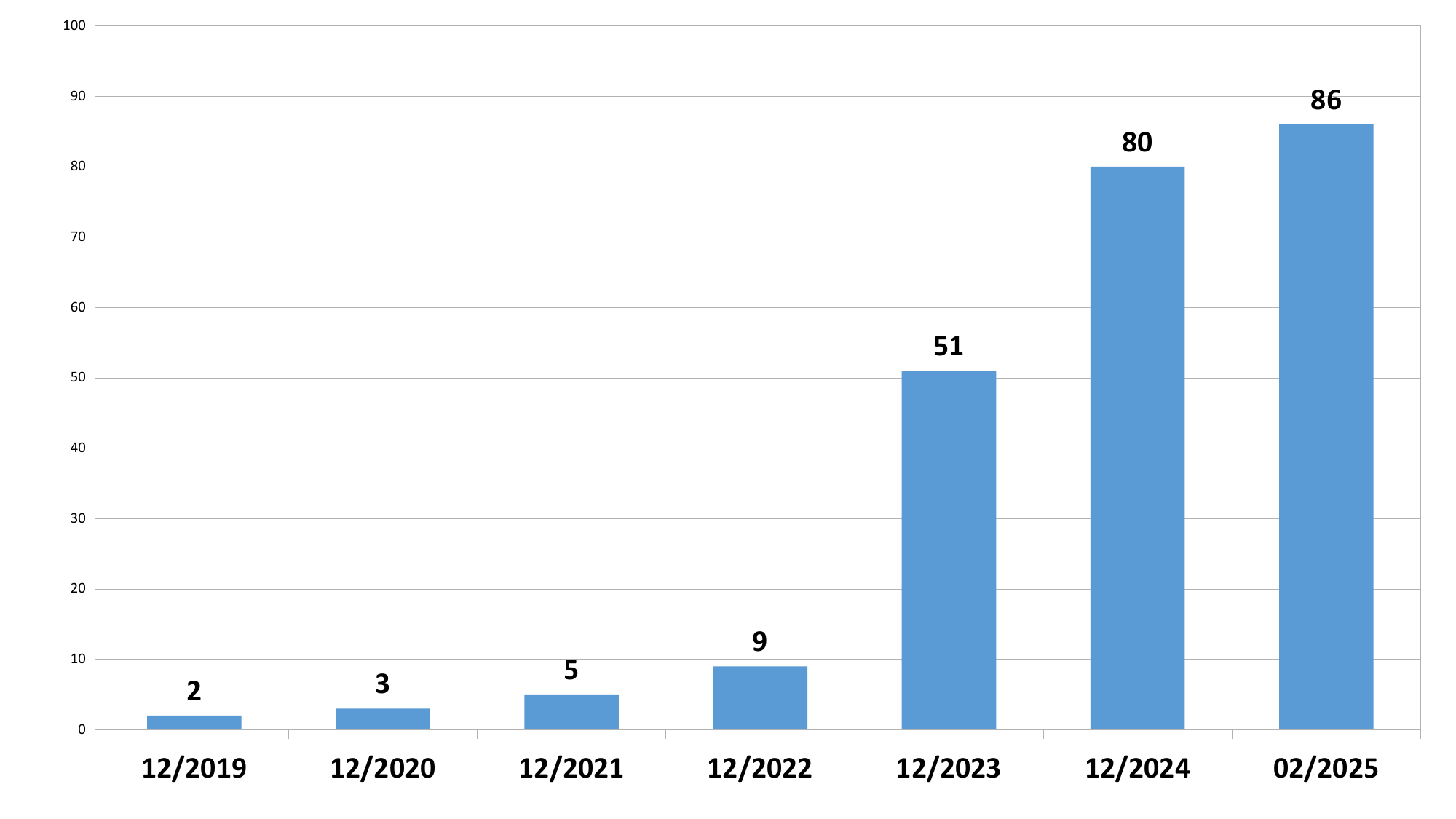}
    \vspace{-2em}
    \caption{Cumulative number of FMs publications over the years}
    \label{fig:year_plot}
\end{figure}

\begin{figure}[t]
    \centering
    \includegraphics[trim=0em 0em 0em 0em, clip, width=\linewidth]{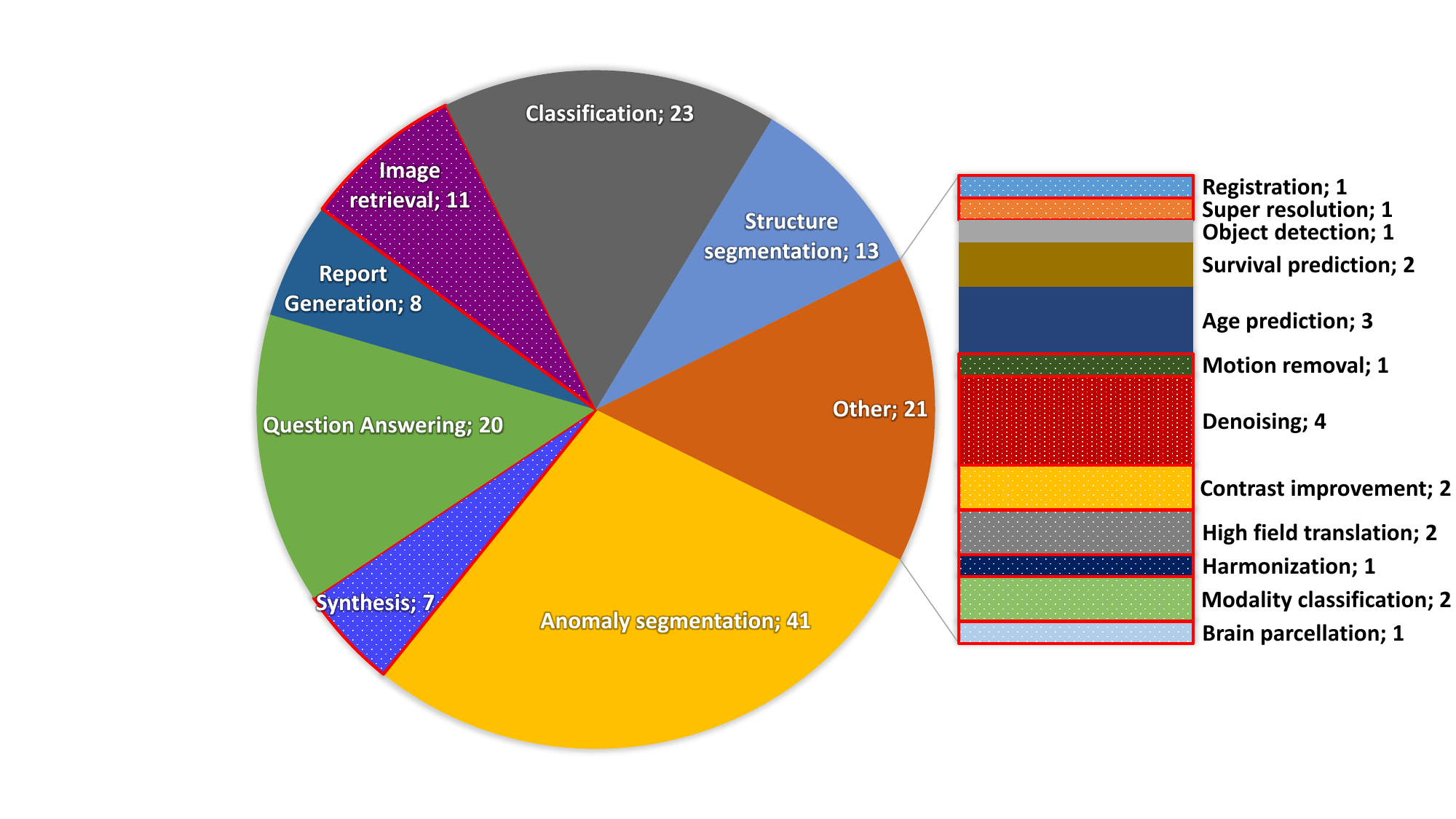}
    \vspace{-2em}
    \caption{Most cited FM over the time period}
    \label{fig:citation_plot}
\end{figure}

\paragraph{Publication timeline}
Our preliminary examination investigates the general trends of research on brain FMs. We present in Fig. \ref{fig:year_plot} the increase in number of publications in the field and in Fig. \ref{fig:citation_plot} the models most cited. Genesis~\cite{zhou_models_2019} and Med3D~\cite{chen_med3d_2019} are the first models that fall under our scope of brain imaging FMs. Although Genesis was published in MICCAI, Med3D was not peer reviewed. Both models use traditional convolution layers and U-Net architectures~\cite{ronneberger2015u}. This year also sees the publication of the first large 3D medical images dataset by the Med3D team, called 3DSeg-8. We compare the models most cited in Section \ref{sec:best-archs} and discuss their main innovations.  

The next two years led to publications with limited citation impact, and there is a significant increase in publications covering brain imaging only in 2023, with 42 publications. 
We believe that this significant increase can be explained by the publication and open-sourcing of very large multipurpose pre-trained FMs such as META's SAM~\cite{kirillov2023segment}. The combination of large imaging datasets (Total Segmentator~\cite{wasserthal2023totalsegmentator}) and pre-trained models is generally a springboard for accelerated research, as was the combination of ImageNet~\cite{deng2009imagenet} and ResNet~\cite{he2016deep} for traditional image classification. 

\paragraph{Publication venue}

\begin{figure}[t]
    \centering
    \includegraphics[trim=7em 0em 7em 0em, clip, width=\linewidth]{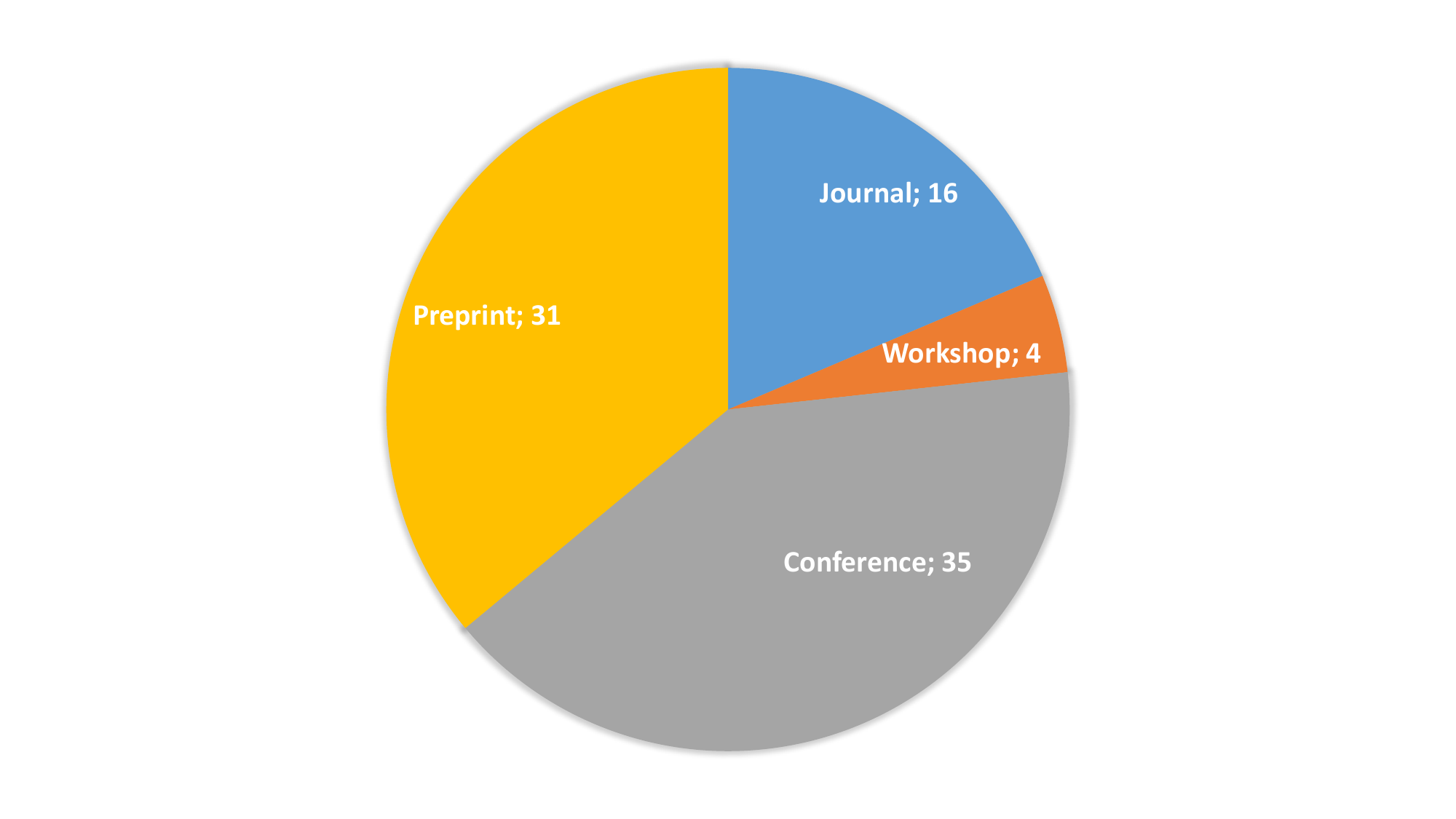}
    \vspace{-2em}
    \caption{Distribution of venues}
    \label{fig:venue_plot}
\end{figure}

We investigate in Fig.\ref{fig:venue_plot} the main venue of publication, and our results show that there are as many publications in peer-reviewed conferences (35) as in the preprint stages (31), and only a minority in journals (16). More notably, the most cited publication in preprint reaches 599 citations (avg 72.13, median 8), while the most cited publication in a conference tops at 698 citations (avg 84.03, median 37.5), and the most cited publication in a journal reaches 1433 citations (avg 131.88, median 11). The relative similarity of metrics between venues hints that very high impact in terms of citations can be achieved undifferentiable through journals, conferences, and preprints, which suggests the fast pace of research in the field and its dynamism. The most common venue of publication for conferences is MICCAI (10), and the most common for journals is IEEE Transactions on Medical Imaging (3). The most cited pre-print is related to the Med3D model, the most cited journal is about the MedSAM model~\cite{ma_medsam_2024} in Nature Communication, and the most cited conference paper introduces LLaVA-Med~\cite{li_llava-med_2023} in NeurIPS.

\begin{figure}[t]
    \centering
    \includegraphics[trim=0em 0em 0em 0em, clip, width=\linewidth]{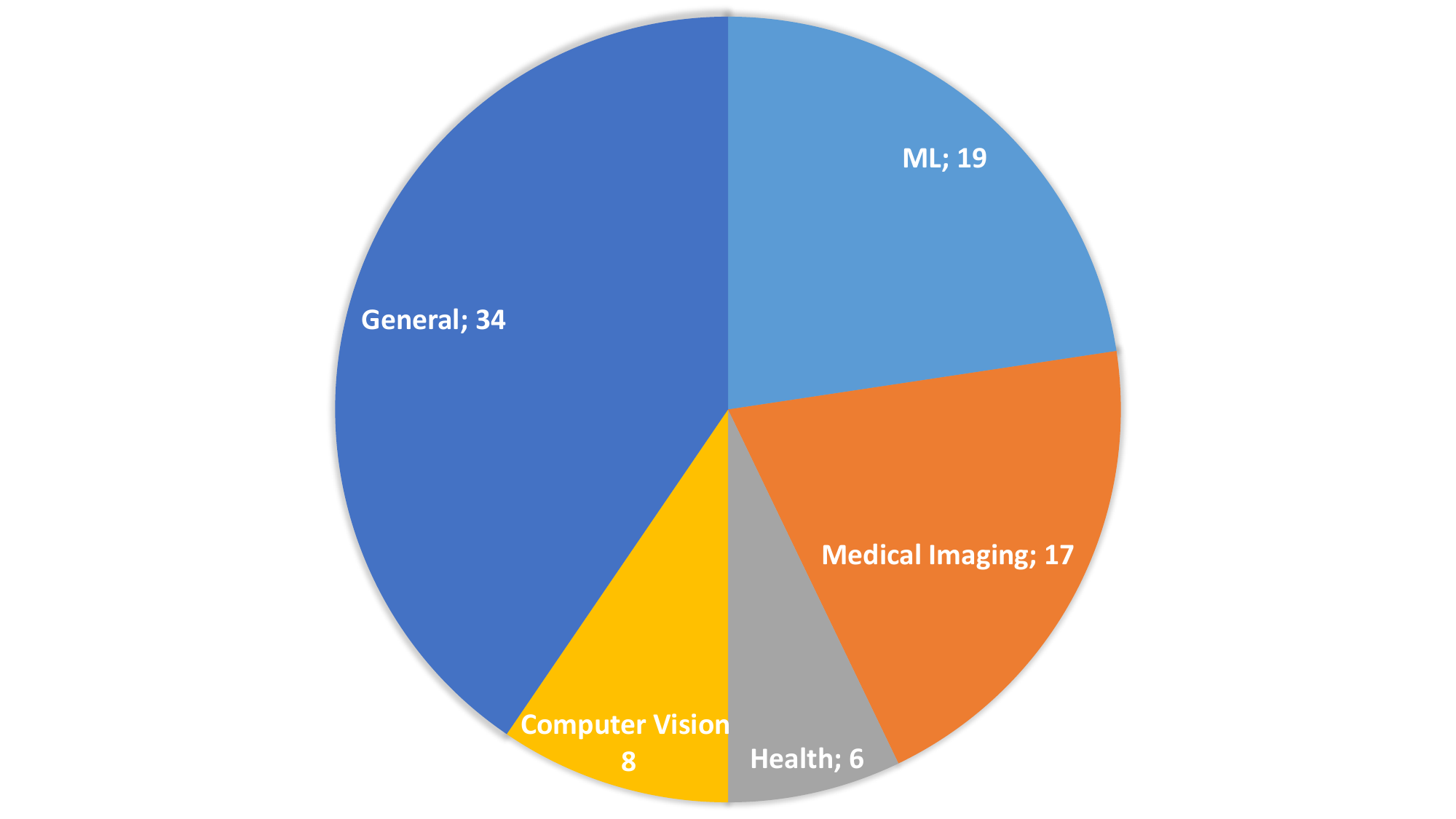}
    \vspace{-2em}
    \caption{Distribution of communities}
    \label{fig:community_plot}
\end{figure}

In general, as Fig.\ref{fig:community_plot} shows, FM research in brain imaging is carried out equivalently in the AI community (machine learning + computer vision) as in the health and medical imaging community, with both 30\% of the publications. FMs for brain imaging are truly cross-disciplinary fields that address challenges relevant to AI and health professionals.  

\paragraph{Source code}

62 FMs are sharing their source code on Github or Huggingface. For 21, the source code does not mention any license, 18 are under the MIT license~\cite{mitlicense}, 16 under the Apache 2.0 license~\cite{apache20}, three under the GNU GPL license~\cite{gpl}, and three under Creative Commons Attribution Noncommercial licenses~\cite{ccby-nc}. Lava-Med is under the Microsoft Research License that restricts commercial applications. 

\begin{Insight}[boxed title style={colback=pink}]
Brain FM research is a fast-paced field that has seen a burst of interest since 2023 in the health and AI communities. Most models disclose their implementation, and notable approaches include Med3D, MedSAM, and LLaVA-Med. 
\end{Insight}

\subsection{Architectures}

\paragraph{Input of the models}
The majority of brain imaging FMs do not support text input (63\%), and are evenly divided between models designed for 2D image inputs (57\%) and 3D image inputs (51\%). Five FMs support 2D and 3D inputs~\cite{wu_radfm_2023,florin_c_ghesu_clopt_2022,archit_medicosam_2025,koleilat_medclip-samv2_2025,nguyen_lvm-med_nodate}. Moreover, 11 FMs support multi-sequence inputs with missing modalities (4D)~\cite{wu_radfm_2023,lei_unibrain_2023,xiaoyu_shi_multimodalsam_2023,cox_brainsegfounder_2024,chen_mengyao_lamim_2024,cecilia_diana-albelda_med-sam-brain_2024,gao_ronghui_pee-fm_2025,xiaoliang_lei_mmc-adapt_2024,tak_divyanshu_brainiac_2024,zhang_mome_2025,g_zhang_brainfound_2025}.

\paragraph{Encoders}
Focusing on visual encoders, 14\% are built on top of CLIP~\cite{eslami_pubmedclip_2023,zhou_anomalyclip_2024,liu_clipgpt_2023,greenspan_oe-vqa_2023,liu_vqa-adapter_2024,cao_dcpl_2024,taha_koleilat_medclip-sam_2024,wang_jinzhuo_minim_2024,koleilat_medclip-samv2_2025,zhu_he_clip-vqg_2023,han_xu_clip-t2medi_2025,wei_xiaoyang_kl-cvr_2024}, 28\% on top of SAM, 23\% on top of U-NET. 14\% of FMs have visual encoders with ResNets architectures (convolution)~\cite{chen_med3d_2019,zhang_medvint_2024,lin_pmc-clip_2023,khare_mmbert_2021,florin_c_ghesu_clopt_2022,nguyen_lvm-med_nodate,liu_t3d_2025,lei_unibrain_2023,carlo_alberto_barbano_anatcl_2024,tak_divyanshu_brainiac_2024,wang_tumsyn_2024,deng_smoe_2025}, and 13\% of FMs designed their encoders using the ViT architecture~\cite{zhang_biomedgpt_nodate,zhang2023biomedclip,wu_radfm_2023,chen_m3ae_2022,chen_med-vlp_2022,liu_qilin-med-vl_2023,chen_ptunifier_2023,greenspan_mumc_2023,du_segvol_2024,chen_mengyao_lamim_2024,gao_ronghui_pee-fm_2025,lee_pirta_2024}. 

Moving to textual encoders and decoders, we also find that a handful of approaches are commonly used as backbones for the models, and some are even mixed. Among the 32 vision-language brain FMs included in our study, 34\% are based on CLIP~\cite{zhou_anomalyclip_2024,eslami_pubmedclip_2023,liu_clipgpt_2023,du_segvol_2024,cao_dcpl_2024,teng_kgpl_2024,taha_koleilat_medclip-sam_2024,koleilat_medclip-samv2_2025}, and 44\% are based on BERT~\cite{zhang_medvint_2024,zhang2023biomedclip,lin_pmc-clip_2023,khare_mmbert_2021,chen_m3ae_2022,zhang_biomedgpt_nodate,qin_miu-vl_2023,chen_med-vlp_2022,greenspan_mumc_2023,chen_ptunifier_2023}. 13\% are built with a LLaMA text encoder~\cite{zhang_medvint_2024,wu_radfm_2023,liu_qilin-med-vl_2023,lee_pirta_2024}, and 16\% of FMs use GPT variants~\cite{chen_medblip_2023,greenspan_oe-vqa_2023,zhu_he_clip-vqg_2023,liu_clipgpt_2023,zhang_biomedgpt_nodate}. 

\paragraph{Other architecture designs}

One brain FM built an architecture that supports RAG~\cite{lee_pirta_2024}, two models support federated learning~\cite{liu_fedfms_2024,asokan_flap-sam_2024}, two models support interactive segmentation~\cite{wong_scribbleprompt_2024,archit_medicosam_2025}, and one model interpretability~\cite{liu_clipgpt_2023}.

\begin{Insight}[boxed title style={colback=pink}]
Although there is a large variety of brain FMs, a few backbones constitute the majority of architectures. Today's FMs are heavily dependant on U-Net, SAM or CLIP visual encoders, and CLIP or BERT text encoders.
\end{Insight}

\subsection{Training}
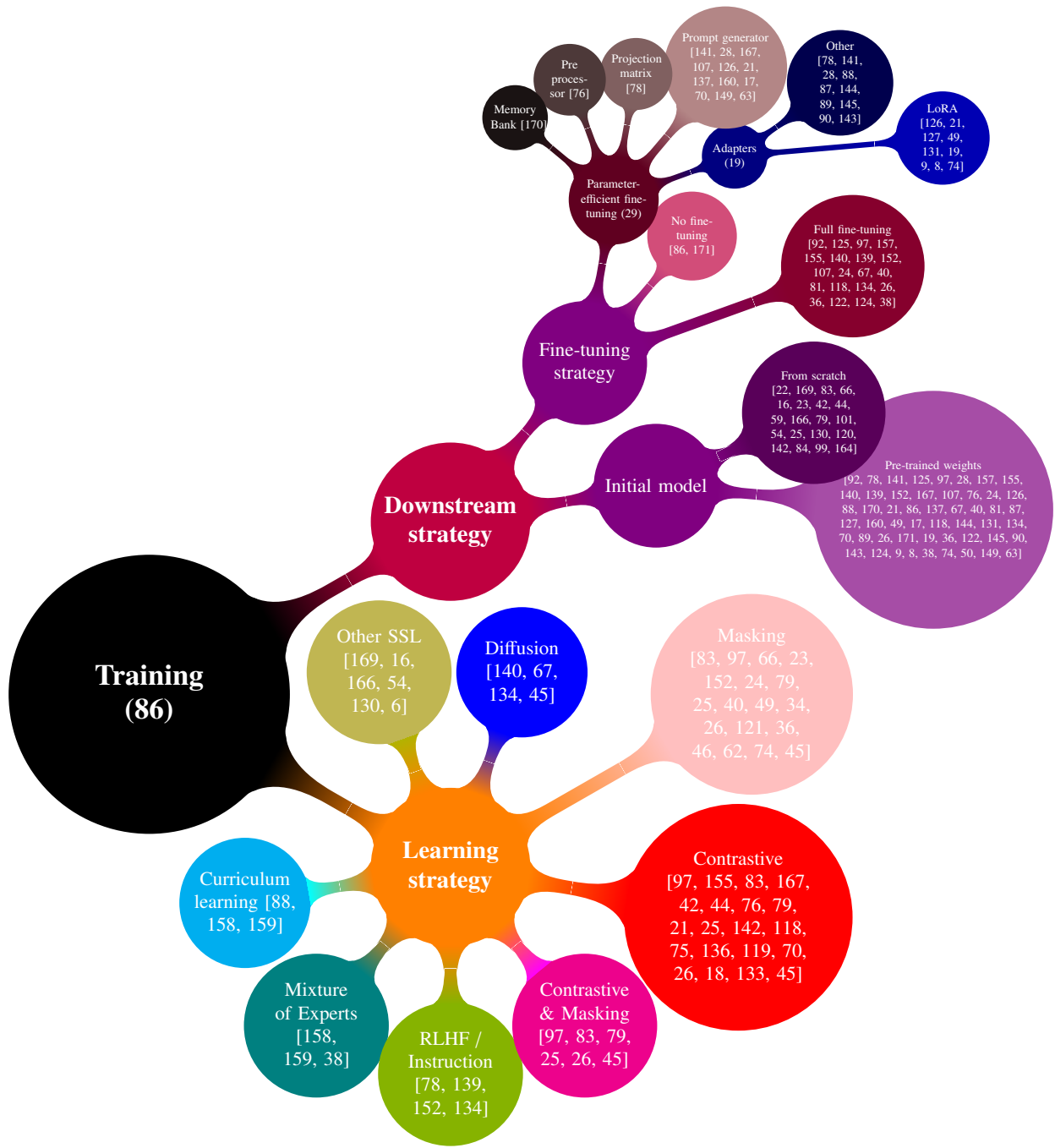
\begin{figure*}[!ht]
\centering
\resizebox{\linewidth}{!}{
\begin{tikzpicture}[
    mindmap,
    grow cyclic,
    every node/.style={concept, text=white, align=center},
    root concept/.append style={
        concept color=black, font=\large\bfseries, text width=2cm},
    level 1 concept/.append style={sibling angle=60, level distance=5cm, font=\bfseries},
    level 2 concept/.append style={sibling angle=40, level distance=3cm},
]

\node[root concept] {Training (86)}
    child[concept color=orange] {
        node {Learning strategy}
        child[concept color=cyan] {node {Curriculum learning \cite{liu_qilin-med-vl_2023, zhang_mome_2024,zhang_mome_2025}}}
        child[concept color=teal] {node {Mixture of Experts \cite{zhang_mome_2024,zhang_mome_2025,deng_smoe_2025}}}
        child[concept color=lime!70!black] {node {RLHF / Instruction \cite{li_llava-med_2023, wu_radfm_2023, zhang_biomedgpt_nodate,wang_jinzhuo_minim_2024}}}
        child[concept color=magenta] {node {Contrastive \& Masking \cite{moor_med-flamingo_2023,lin_pmc-clip_2023,greenspan_mumc_2023,chen_ptunifier_2023,chen_mengyao_lamim_2024,g_zhang_brainfound_2025}}}
        child[concept color=red, level distance=4.2cm, text width=2.5cm] {node {Contrastive \\ \cite{moor_med-flamingo_2023,zhang_biomedclip_2025,lin_pmc-clip_2023,zhou_anomalyclip_2024,eslami_pubmedclip_2023,florin_c_ghesu_clopt_2022,lei_medlsam_2024,greenspan_mumc_2023,chen_medblip_2023,chen_ptunifier_2023,wu_voco_2024,taha_koleilat_medclip-sam_2024,lei_unibrain_2023,wei_xiaoyang_kl-cvr_2024,tak_divyanshu_brainiac_2024,koleilat_medclip-samv2_2025,chen_mengyao_lamim_2024,carlo_alberto_barbano_anatcl_2024,wang_tumsyn_2024,g_zhang_brainfound_2025}}}
        child[concept color=pink, level distance=5cm, text width=2cm] {node {Masking \\ \cite{lin_pmc-clip_2023,moor_med-flamingo_2023,khare_mmbert_2021,chen_m3ae_2022,zhang_biomedgpt_nodate,chen_med-vlp_2022,greenspan_mumc_2023,chen_ptunifier_2023,du_segvol_2024,gu_finetune-sam_2024,cox_brainsegfounder_2024,chen_mengyao_lamim_2024,tassilo_wald_mae-seg_2024,d_wood_neurobert_2024,gao_ronghui_pee-fm_2025,joseph_cox_brainfounder_2024,lee_pirta_2024,g_zhang_brainfound_2025}}}
        child[concept color=blue] {node {Diffusion \cite{wu_medsegdiff-v2_2023,kim_jonghun_aldm_2024,wang_jinzhuo_minim_2024,g_zhang_brainfound_2025}}}
        child[concept color=yellow!70!black] {node {Other SSL \cite{zhou_models_2019,butoi_universeg_2023,zhou_pcrlv2_2023,he_gvsl_2023,wang_mis-fm_2023,alan_q_wang_brainmorph_2024}}}
    }
    child[concept color=purple] {
        node {Downstream strategy}
        child[concept color=violet] {
            node {Initial model}
            child[concept color=violet!70!white, level distance=4cm, text width=3cm] {node { Pre-trained weights \\ \cite{ma_medsam_2024,li_llava-med_2023,wu_med-sa_2023,tu_med-palmm_2023,moor_med-flamingo_2023,cheng_sam-med2d_2023,zhang_medvint_2024,zhang_biomedclip_2025,wu_medsegdiff-v2_2023,wu_radfm_2023,zhang_biomedgpt_nodate,zhou_anomalyclip_2024,qin_miu-vl_2023,lei_medlsam_2024,chen_med-vlp_2022,greenspan_oe-vqa_2023,liu_qilin-med-vl_2023,zhu_medical_2024,chen_medblip_2023,liu_clipgpt_2023,wong_scribbleprompt_2024,kim_jonghun_aldm_2024,du_segvol_2024,li_yunxiang_nnsam_2024,liu_vqa-adapter_2024,w_feng_eff-sam_2023,zhang_ur-sam_2024,gu_finetune-sam_2024,cao_dcpl_2024,taha_koleilat_medclip-sam_2024,xiaoyu_shi_multimodalsam_2023,wang_sam-med3d_2024,wang_jinzhuo_minim_2024,koleilat_medclip-samv2_2025,liu_fedfms_2024,chen_mengyao_lamim_2024,zhu_he_clip-vqg_2023,cecilia_diana-albelda_med-sam-brain_2024,d_wood_neurobert_2024,teng_kgpl_2024,xiaoyu_shi_medicalsam_2024,liu_progressivead_2024,xiaoliang_lei_mmc-adapt_2024,tobari_shuya_se-ada_2025,asokan_flap-sam_2024,archit_medicosam_2025,deng_smoe_2025,lee_pirta_2024,han_xu_clip-t2medi_2025,yongeun_jang_ssnet_2024,kai_zhou_obj-sam_2024}}}
            child[concept color=violet!70!black, level distance=2.5cm, text width=1.5cm] {node {From scratch \cite{chen_med3d_2019,zhou_models_2019,lin_pmc-clip_2023,khare_mmbert_2021,butoi_universeg_2023,chen_m3ae_2022,eslami_pubmedclip_2023,florin_c_ghesu_clopt_2022,huang_stu-net_2023,zhou_pcrlv2_2023,greenspan_mumc_2023,nguyen_lvm-med_nodate,he_gvsl_2023,chen_ptunifier_2023,wang_mis-fm_2023,tao_li_mrac_2021,wu_voco_2024,liu_t3d_2025,murmu_anita_3dunettumour_2021,zhong_liming_mtt-net_2024}}}
        }
        child[concept color=violet] {
            node {Fine-tuning strategy}
            child[concept color=purple!70!black, level distance=4.1cm, text width=1.5cm] {node {Full fine-tuning \cite{ma_medsam_2024,tu_med-palmm_2023,moor_med-flamingo_2023,zhang_medvint_2024,zhang_biomedclip_2025,wu_medsegdiff-v2_2023,wu_radfm_2023,zhang_biomedgpt_nodate,qin_miu-vl_2023,chen_med-vlp_2022,kim_jonghun_aldm_2024,du_segvol_2024,li_yunxiang_nnsam_2024,taha_koleilat_medclip-sam_2024,wang_jinzhuo_minim_2024,chen_mengyao_lamim_2024,d_wood_neurobert_2024,teng_kgpl_2024,tobari_shuya_se-ada_2025,deng_smoe_2025}}}
            child[concept color=purple!70!white] {node {No fine-tuning \cite{liu_clipgpt_2023,zhu_he_clip-vqg_2023}}}
            child[concept color=purple!50!black] {
                node {Parameter-efficient fine-tuning (29)}
                child[concept color=blue!50!black] {
                    node {Adapters (19)}
                    child[concept color=blue!70!black, level distance=3cm] {node {LoRA \cite{greenspan_oe-vqa_2023,chen_medblip_2023,w_feng_eff-sam_2023,gu_finetune-sam_2024,wang_sam-med3d_2024,cecilia_diana-albelda_med-sam-brain_2024,asokan_flap-sam_2024,archit_medicosam_2025,lee_pirta_2024}}}
                    child[concept color=blue!30!black] {node {Other \cite{li_llava-med_2023,wu_med-sa_2023,cheng_sam-med2d_2023,liu_qilin-med-vl_2023,liu_vqa-adapter_2024,xiaoyu_shi_multimodalsam_2023,liu_fedfms_2024,xiaoyu_shi_medicalsam_2024,liu_progressivead_2024,xiaoliang_lei_mmc-adapt_2024}}}
                }
                child[concept color=pink!70!black, level distance=2.5cm, text width=1.3cm] {node {Prompt generator \\ \cite{wu_med-sa_2023,cheng_sam-med2d_2023,zhou_anomalyclip_2024,qin_miu-vl_2023,greenspan_oe-vqa_2023,chen_medblip_2023,wong_scribbleprompt_2024,zhang_ur-sam_2024,cao_dcpl_2024,koleilat_medclip-samv2_2025,yongeun_jang_ssnet_2024,kai_zhou_obj-sam_2024}}}
                child[concept color=pink!50!black] {node {Projection matrix ~\cite{li_llava-med_2023}}}
                child[concept color=pink!30!black] {node {Pre processor~\cite{lei_medlsam_2024}}}
                child[concept color=pink!10!black] {node {Memory Bank~\cite{zhu_medical_2024}}}
            }
        }
    };

\end{tikzpicture}
}
\caption{Mindmap of Training Techniques of FMs}
\label{fig:mindmap-training}
\end{figure*}

We summarize the training strategies in Fig.\ref{fig:mindmap-training}. Most models were fine-tuned while freezing a part of the model, 33 with a frozen encoder and 9 with a frozen decoder. 

\paragraph{Learning strategy}

About half of FMs used either Contrastive Learning or Masked Input Learning as self-supervised pre-training strategy. Masked learning was used in the visual encoder for 15\% of models, and in the language encoder for 9\% of the models.  

Other Self-supervised Learning strategies include Image Text Matching \cite{chen_med-vlp_2022,greenspan_mumc_2023} and Graph Matching \cite{chen_med-vlp_2022,nguyen_lvm-med_nodate}.

\paragraph{Downstream strategy}
To train the models for their downstream tasks, a large variety of approaches were used.
Only 41\% of the models were trained from scratch; the remaining FMs started with pre-trained weights. Among these, 57\% used some sort of parameter efficient fine-tuning: 37\% relied on adapter fine-tuning (including Low Rank Adaptation \cite{greenspan_oe-vqa_2023,chen_medblip_2023,w_feng_eff-sam_2023,gu_finetune-sam_2024,wang_sam-med3d_2024,cecilia_diana-albelda_med-sam-brain_2024,asokan_flap-sam_2024,archit_medicosam_2025,lee_pirta_2024}) and 24\% proposed to optimize the prompt generator.  

\begin{Insight}[boxed title style={colback=pink}]
Most brain FMs start from pre-trained weights and use efficient fine-tuning approaches, like adapters. When trained from scratch, brain FMs rely mostly for pre-training on contrastive and masked learning.
\end{Insight}

\subsection{Tasks}

\begin{figure}[t]
    \centering
    \includegraphics[trim=0em 0em 0em 0em, clip, width=\linewidth]{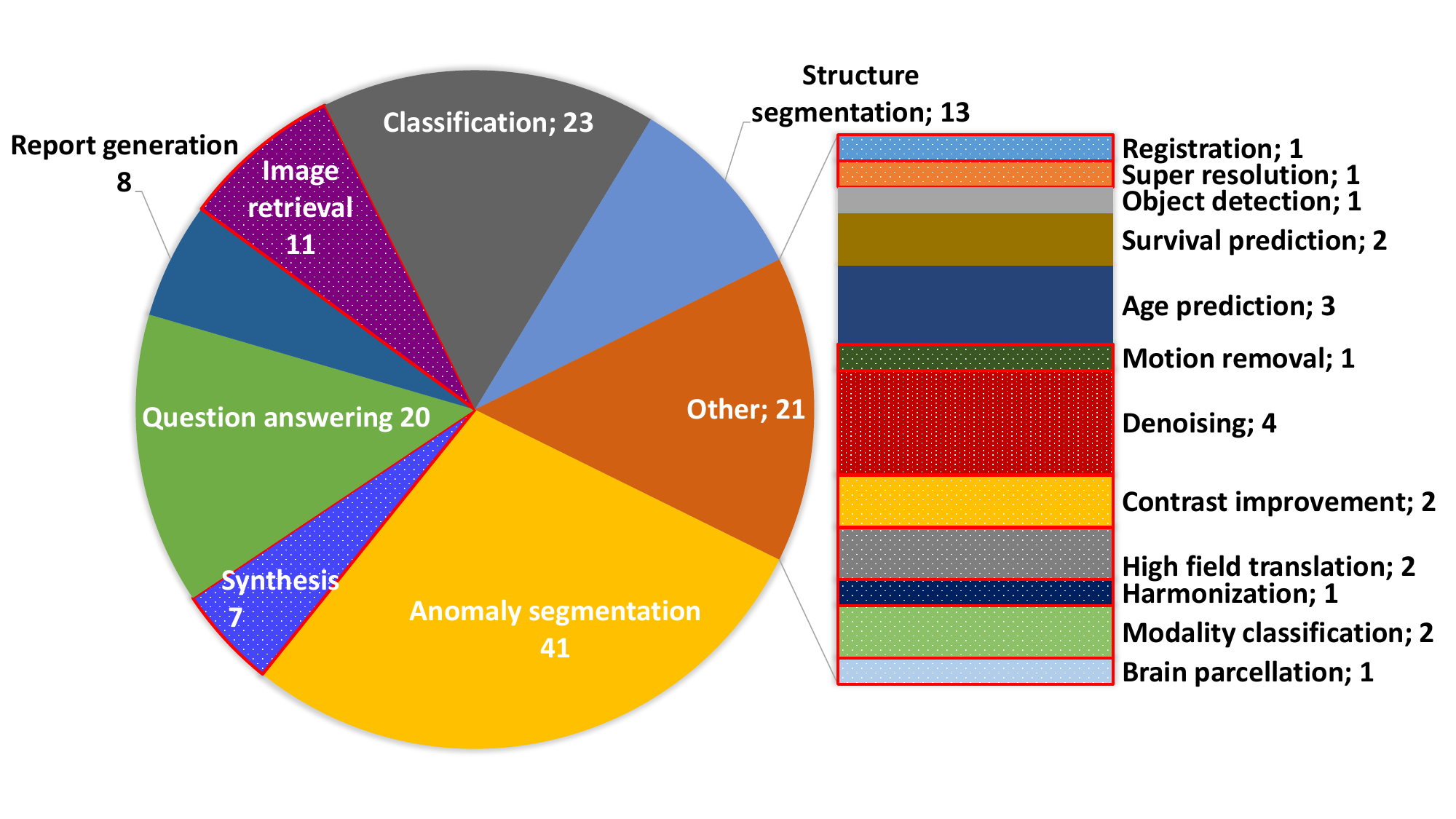}
    \vspace{-2em}
    \caption{Distribution of tasks. The dashed tasks with red borders are auxiliary tasks.}
    \label{fig:tasks_plot}
\end{figure}

In this section, we explore the tasks that the surveyed literature handles and the brain pathologies and datasets used to evaluate these tasks. We show in Fig.~\ref{fig:tasks_plot} the number of models that support each task. The tasks with a red border are preprocessing or auxiliary tasks (i.e., non-diagnosis tasks), while the remaining tasks are diagnosis tasks. 

In total, 28\% of the models provide non-diagnosis tasks; in particular, 13\% support image retrieval~\cite{zhang_biomedclip_2025,lin_pmc-clip_2023,chen_m3ae_2022,chen_med-vlp_2022,chen_ptunifier_2023,liu_t3d_2025,taha_koleilat_medclip-sam_2024,koleilat_medclip-samv2_2025,d_wood_neurobert_2024,wei_xiaoyang_kl-cvr_2024}, and 8\% support image synthesis~\cite{kim_jonghun_aldm_2024,tao_li_mrac_2021,zhong_liming_mtt-net_2024,wang_jinzhuo_minim_2024,wang_tumsyn_2024,g_zhang_brainfound_2025,han_xu_clip-t2medi_2025}. 

90\% of FMs provide diagnosis tasks. The main diagnosis tasks are anomaly segmentation with 48\% of FM models, anomaly classification (27\%) and question-answering (23\%). 

In total, 20\% of FMs learn at least one non-diagnosis task and one diagnosis task. The least common tasks overall with one model each are Image Registration, Super-Resolution, Object Detection, Motion Removal, Harmonization, and Brain Parcellation.

\begin{figure}[t]
    \centering
    \includegraphics[trim=5em 0em 5em 0em, clip, width=\linewidth]{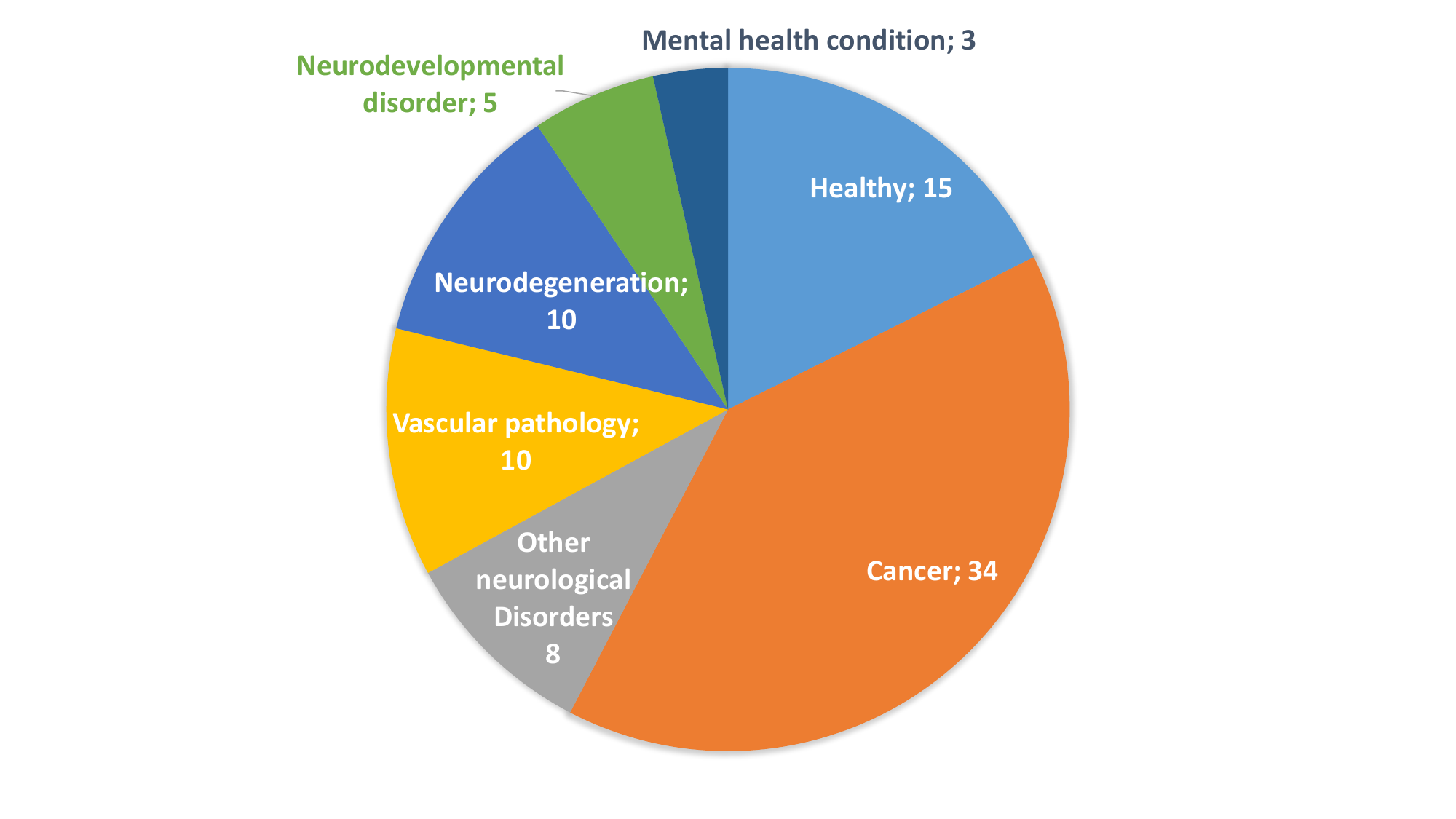}
    \vspace{-2em}
    \caption{Distribution of pathologies in the Foundation Models.}
    \label{fig:pathologies_plot}
\end{figure}

\subsection{Datasets}

We show in Fig.~\ref{fig:pathologies_plot} the distribution of the pathologies covered in the FM. Most FMs tackled brain cancer (40\%), followed by healthy brains (17\%)~\cite{wu_med-sa_2023,kim_jonghun_aldm_2024,wood_david_a_brainage_2024,cox_brainsegfounder_2024,tassilo_wald_mae-seg_2024,zhang_mome_2024,zhang_mome_2025,alan_q_wang_brainmorph_2024,sun_yue_bme-x_2024,wang_tumsyn_2024,joseph_cox_brainfounder_2024,asokan_flap-sam_2024,tak_divyanshu_brainiac_2024,deng_smoe_2025,g_zhang_brainfound_2025}. In addition, 63\% of brain FM we trained with a combination of brain images and datasets related to other organs such as the chest, kidney, and liver. About 35\% FMs use brain image datasets with unclear or inexistent pathological annotations (for example, datasets scrapped from PubMed articles with coarse labeling, such as brain MRI / CT, instead of the exact type of pathology).   

When divided by imaging modality, 88\% of the FMs cover MRI images, 62\% used CT images, and only 6\% support PET images\cite{cheng_sam-med2d_2023,zhang2023biomedclip,huang_stu-net_2023,tao_li_mrac_2021,liu_progressivead_2024}. In total, 78\% of FMs were trained on more than one type of modality.

\begin{Insight}[boxed title style={colback=pink}]
There is a large imbalance in the pathologies and tasks faced by brain FMs. Although MRI, cancer, and segmentation have been very well explored, niches such as PET imaging and mental health conditions remain underexplored.  
\end{Insight}

\section{Findings: Brain Imaging Datasets}
\label{sec-datasets}

To evaluate whether the research on brain FMs matches the existing literature on brain imaging, we collected a list of 146 3D brain imaging datasets and 15 2D brain imaging datasets. We relied on FMs reference citations and surveying the major brain imaging repositories: TCIA\footnote{\url{https://www.cancerimagingarchive}}, OpenNeuron\footnote{\url{https://openneuro.org}}, and Synapse\footnote{\url{https://www.synapse.org}}.

We will first present the main 2D imaging datasets with brain images, then explore in more detail the 3D brain imaging datasets. Indeed, brain imaging is originally volumetric, and the 2D datasets only use slices extracted from publications figures, or from the 3D volumes, and do not represent original sources. 

In the following, when providing the number of studies (3d images) for a data set with multiple pathological labels (eg, healthy, high-grade glioma and low-grade glioma), we consider the \textbf{total number of studies} in all of them. 

We provide an interactive atlas to explore the datasets on the link \url{https://ballistic-purple-6f5.notion.site/176bf710ca5980d2bccdd6029523f3ed?v=a9f1e6e135d84abdac7764776353dd7c}. It allows multi-criteria filtering and provides the links to request or download the datasets from their respective sources.

\subsection{2D Brain images.}

Datasets of 2D brain images are mostly scraped from large repositories. They are generally extracted in pairs with their caption for visual questions-answer tasks or classification. Given the automated scrapping and parsing of these datasets, they generally mix different imaging modalities (MRI, CT, PET) and sequences (T1W, T2W, etc.) with captions of varying detail levels. We report in Table \ref{tab:2d_datasets} the main 2D Imaging datasets with their citation counts in Google Scholar. Some datasets were published along models, and the count may reflect the interest of both the dataset and the model (on 01/04/2025). 

9/15 of the 2d imaging datasets are multimodal datasets, with medical images and associated text data: figure captions, clinical reports, engineered pairs for Question Answering (QA) or specific usage pairs (CLIP). The remaining six datasets are image-only, but provide masks, either for tumors and lesions, or mixed segmentation masks (organs, tissues, tumors, ...) depending on the context \cite{ye2023sa}. The most cited multimodal dataset is ROCO with 504 citations combining the two versions, and the most cited image-only dataset is COSMOS 1050K with 375 citations. The multimodal datasets are all mixed sequences, mixing MRI, CT, PET and other protocols not covered (X-ray, ultrasound, etc.), while the six image-only datasets contain only one or two types of sequence each (except the general purpose dataset SA-Med2D-20M \cite{ye2023sa}), five with MRI, four with CT, and three with PET.

The vast majority of datasets cover multiple organs, including brains (11 datasets designated "General"), one data set focuses on multiple sclerosis\cite{muslim2022brain}, one on multiple brain pathologies \cite{zhang2025multimodal}, and the remaining three are brain cancer datasets.  

\begin{Insight}[boxed title style={colback=pink}]
Although general image datasets combining scrapped images and annotations collected a large amount of images (including brain data), specialized 2D brain datasets remain small for some imaging modalities (MRI, PET).   
\end{Insight}

\begin{table}[t]
    \centering
\resizebox{\linewidth}{!}{
    \begin{tabular}{l|l|l|l|l|l}

    \textbf{Name} & \textbf{Images} & \textbf{Type} & \textbf{Annotations} & \textbf{Pathologies} & \textbf{\# citations} \\
    \hline
    MTB \cite{moor2023med}& 25M+ & PET/CT/MRI & Image-text & General &  255\\
    AutoPET \cite{gatidis2022whole} & 1,014 & PET/CT & Tumor masks & General & 238 \\
    HECKTOR \cite{oreiller2022head} & 883 & PET/CT & Tumour masks & Head/neck cancer &  221\\
    MRI-Scler \cite{muslim2022brain}& 42 & MRI & Lesion masks & Multiple sclerosis &  28 \\
    ImageClef-VQA-2019 \cite{benabacha2019vqamed}& 4,500 & Mixed & QA pairs & General & 249 \\
    BrainCT-3M \cite{zhang2025multimodal} & 3M+ & CT & Clinical reports & Brain diseases & 0\\
    PMC-VQA \cite{zhang2023pmc}& 15M+ & Mixed & QA pairs & General &  200\\
    COSMOS 1050K \cite{huang2024segment}& 1.05M & CT/MRI & Segmentation & General & 375 \\
    MedICaT \cite{subramanian2020medicat}& 217K+ & Mixed & Captions & General & 84 \\
    ROCO \cite{pelka2018roco}& 44,084 & CT/MRI & Captions & General & 223\\
    ROCOV2 \cite{ruckert2024rocov2}& 79,789 & Mixed & Captions & General & 281\\
    BrainTumourDataset\cite{cheng2017brain} & 3,064 & MRI & Tumor masks & Brain tumors & 246 \\
    SA-Med2D-20M  \cite{ye2023sa}& 20M & Mixed & Segmentation & General & 32 \\
    PMC-15M \cite{zhang2023biomedclip}& 15M & Mixed & Image-text & General & 211 \\
    PMC-OA \cite{lin2023pmc}& 1.65M & Mixed & CLIP pairs & General &  211 \\
    CCBTM \cite{hashemi2023crystal} & 21,672 & MRI & Tumor masks & Brain tumors & 14 \\

    \end{tabular}
}
    \caption{2D Brain Imaging datasets}
    \label{tab:2d_datasets}
\end{table}

\subsection{3D Brain images.}

\paragraph{Dataset deduplication}
The 146 identified datasets represent 541,042 studies (3D images) in all pathologies and sequence types. 18 datasets (for 34,378 images) are clearly duplicates or variants of other datasets.

\begin{Insight}[boxed title style={colback=pink}]
About 6\% of the 3D medical imaging studies are from duplicate or derivative datasets, which poses serious risks of data leakage when mixing various datasets for one model training. 
\end{Insight}

\paragraph{Pathologies}

Following the pathology nomenclature for brain FMs, we sort the datasets according to the type of pathologies covered. There are eight datasets for vascular pathologies (46,081 studies), 53 datasets for brain cancer (39,101 studies), nine datasets for neurological disorders (41,720 studies), 17 datasets (66,070 studies) for neurodegenerative pathologies, three datasets (59,162 studies) for neurodevelopmental disorders, nine datasets for mental health conditions (31,585 studies), one dataset for HIV infections (351 studies), and two datasets for intoxications such as drug abuse (100 studies) and alcoolism (1,182 studies).

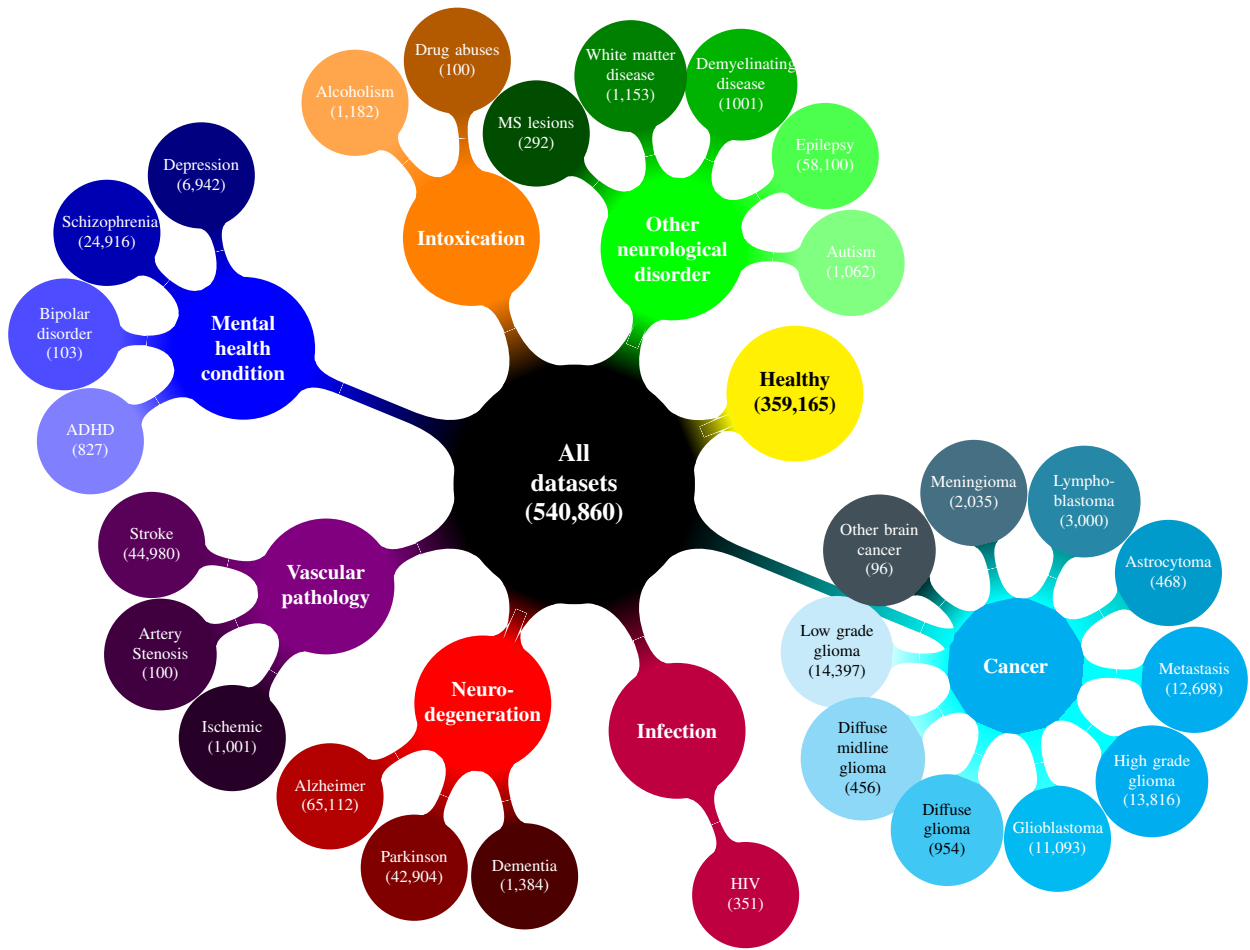
\begin{figure*}[ht]
\centering
\resizebox{\linewidth}{!}{
\begin{tikzpicture}[
    mindmap,
    grow cyclic,
    every node/.style={concept, text=white, align=center},
    root concept/.append style={
        concept color=black, font=\large\bfseries, text width=2cm},
    level 1 concept/.append style={sibling angle=45, level distance=4.5cm, font=\bfseries},
    level 2 concept/.append style={sibling angle=36, level distance=3cm},
]

\node[root concept] {All datasets \\ (540,860)}
    child[concept color=violet, level distance=4.5cm] { node {Vascular pathology}
        child[concept color=violet!70!black] { node {Stroke \\ (44,980)} }
        child[concept color=violet!50!black] { node {Artery Stenosis \\ (100)} }
        child[concept color=violet!30!black] { node {Ischemic \\ (1,001)} }
    }
     child[concept color=red, level distance=4cm]{ node {Neuro-degeneration}
        child[concept color=red!70!black] { node {Alzheimer \\ (65,112)} }
        child[concept color=red!50!black] { node {Parkinson \\ (42,904)} }
        child[concept color=red!30!black] { node {Dementia \\ (1,384)} }
    }
    child[concept color=purple]{ node {Infection}
        child[concept color=purple]{ node {HIV \\ (351)} }
    }
    child[concept color=cyan, level distance=8cm]{ node {Cancer}
        child[concept color=cyan!20!white]{ node[text=black] {Low grade glioma \\ (14,397)} }
        child[concept color=cyan!40!white]{ node[text=black] {Diffuse midline glioma \\ (456)} }
        child[concept color=cyan!60!white]{ node[text=black] {Diffuse glioma \\ (954)} }
        child[concept color=cyan!80!white]{ node[] {Glioblastoma \\ (11,093)} }
        child[concept color=cyan]{ node[] {High grade glioma \\ (13,816)} }
        child[concept color=cyan]{ node[] {Metastasis \\ (12,698)} }
        child[concept color=cyan!80!black]{ node[] {Astrocytoma \\ (468)} }
        child[concept color=cyan!60!black]{ node[] {Lympho-blastoma \\ (3,000)} }
        child[concept color=cyan!40!black]{ node[] {Meningioma \\ (2,035)} }
        child[concept color=cyan!20!black]{ node[] {Other brain cancer \\ (96)} }
    }
    child[concept color=yellow, level distance=4cm] { node[text=black] {Healthy (359,165)} }
    child[concept color=green, level distance=4.3cm]{ node {Other neurological disorder}
        child[concept color=green!50!white]{ node {Autism \\ (1,062)} }
        child[concept color=green!70!white]{ node {Epilepsy \\ (58,100)} }
        child[concept color=green!70!black]{ node {Demyelinating disease \\ (1001)} }
        child[concept color=green!50!black]{ node {White matter disease \\ (1,153)} }
        child[concept color=green!30!black]{ node {MS lesions \\ (292)} }
    }
    child[concept color=orange]{ node {Intoxication}
        child[concept color=orange!70!black]{ node {Drug abuses \\ (100)} }
        child[concept color=orange!70!white]{ node {Alcoholism \\ (1,182)} }
    }
    child[concept color=blue, level distance=6cm]{ node {Mental health condition}
        child[concept color=blue!50!black]{ node {Depression \\ (6,942)} }
        child[concept color=blue!70!black]{ node {Schizophrenia \\ (24,916)} }
        child[concept color=blue!70!white]{ node {Bipolar disorder \\ (103)} }
        child[concept color=blue!50!white]{ node {ADHD \\ (827)} }
    }
    ;
\end{tikzpicture}
}
\caption{Mindmap of 3D Brain MRI datasets split by pathologies. In parentheses the total number of studies of the datasets}
\label{fig:mindmap-3d-datasets}
\end{figure*}

In Figure \ref{fig:mindmap-3d-datasets} we provide a detailed nomenclature of datasets and the count of studies per pathology. Alzheimer (65,112 studies),  epilepsy (58,100 studies), and brain stroke (44,980) are the brain pathologies with the most available data. Low-grade glioma with 14,397 studies across 19 datasets is the most common cancer in brain medical imaging datasets. It is worth noting that some pathology labels can be subsumed. For example, MS lesions are a demyelating disease, and gliobastoma are a subset of high grade cancers. Meanwhile, some pathologies can overlap, a dataset of Astrocytoma can contain low grade and high grade astrocytoma. 

Eight datasets have collected longitudinal data for all or a subset of their cohort (3 cancer, 2 alzheimer, 3 healthy).

\begin{Insight}[boxed title style={colback=pink}]
There is a huge unbalanced in medical imaging data availability across brain pathologies. In addition, the label categories of some datasets can subsume or overlap the categories from another dataset.   
\end{Insight}

\paragraph{Modalities}

\begin{figure}[H]
    \centering
    \includegraphics[trim=0em 0em 0em 0em, clip, width=\linewidth]{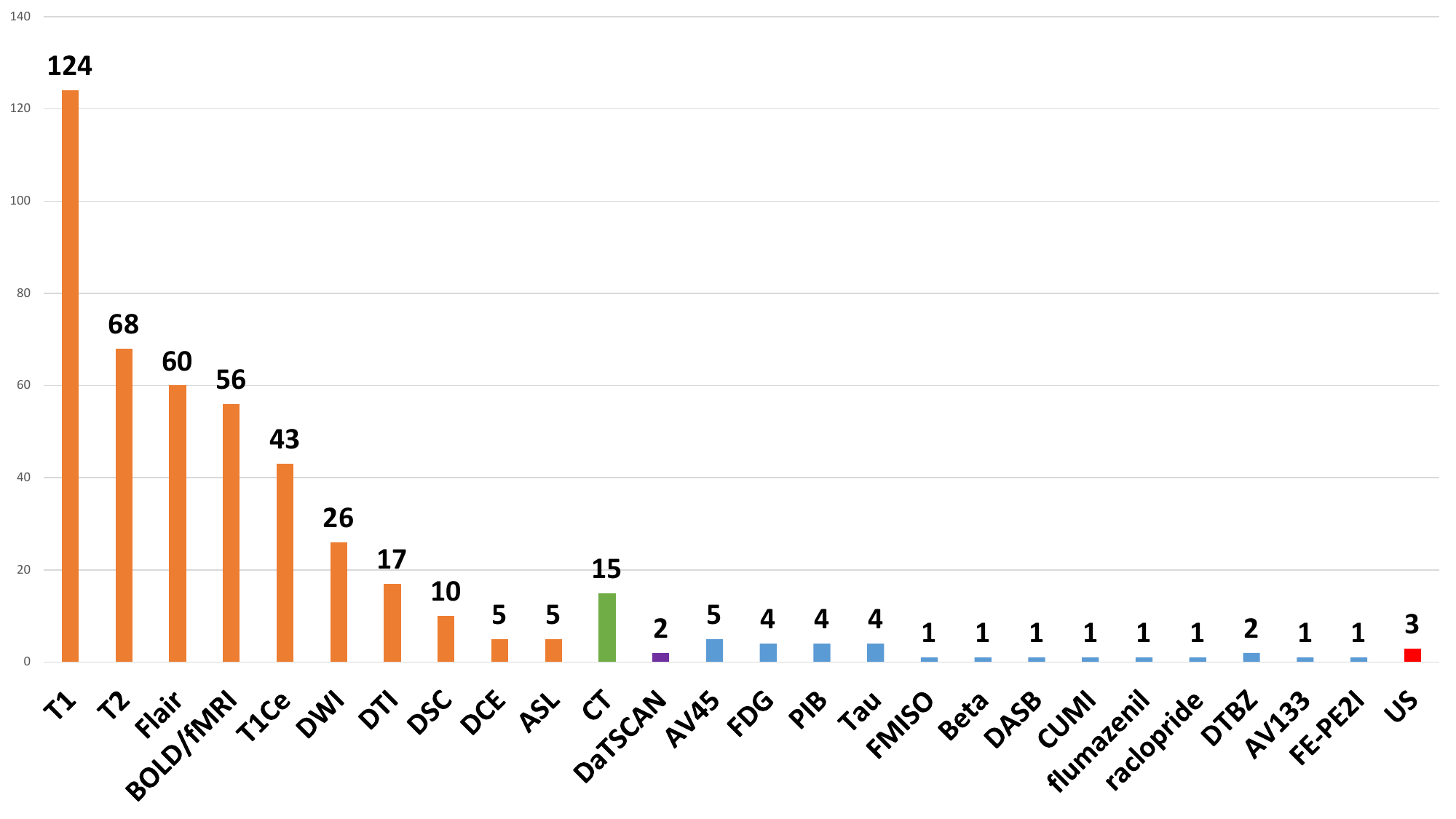}
    \vspace{-2em}
    \caption{Number of datasets per imaging modality and sequences. In orange MRI sequences, in green CT, in purple SPECT, in blue PET tracers and in red ultrasound.}
    \label{fig:dataset_modalities}
\end{figure}

We compare in Figure \ref{fig:dataset_modalities} the number of datasets providing each imaging modality. While we identified three datasets with ultra-sound and two datasets of SPECT, the majority of datasets covers PET (13 datasets), CT (15 datasets), and MRI (134 datasets). 

In MRI imaging, T1 (mostly T1w) sequences are the most available (124 datasets representing 512,106 studies), followed by T2w (68 datasets, 350,452 studies), FLAIR (60 datasets, 127,028 studies) and BOLD/fMRI (56 datasets, 387,807 studies). The high number of studies per datasets for BOLD/fMRI is explained by the fact that fMRI data is collected per patient and per task, and multiple tasks are typically collected per patients, leading to multiple studies for the same acquisition session.

PET imaging has a variety of tracers collected across different datasets, and multiple tracers can be collected for the same patient in a cohort, for example, to quantify amyloid plaques that characterize Alzheimer's disease. 

In addition to the original imaging contrast / tracer data, 38 datasets provided segmentation masks for tumours or lesions and 17 provided structure segmentations of the brain. 

\begin{Insight}[boxed title style={colback=pink}]
While common MRI sequences such as T1, T2 and FLAIR are largely available in the published datasets, datasets with rarer MRI sequences such as DCE, and PET tracers remain scarce. 
\end{Insight}

\paragraph{Access}

Access to high quality public and license-friendly datasets is a cornerstone to the development of large FMs, and our analysis of the existing datasets show a contrasted, but understandable setting: Only 46 datasets among the 146 are publicly available for download. This number jumps to 108 if we consider datasets that require accepting a specific data user agreement for a specific purpose. It is then up to the dataset provider to accept or not the usage of the data to train a multi-purpose FM. Indeed, data privacy laws usually require an anonymization of the patients' data. 36 datasets have explicitly mentioned an anomymization process, including tasks such as skull-stripping (8) or defacing (11).  Another constraint to the democratization of these datasets for brain FMs is licensing schemes. 82 use a license allowing a commercial application, 16 explicitely forbid a commercial use of the datasets, and 48 do not mention explicitly a license. 

The data capture and access also varies depending on which inhabitants took part in the cohorts. For 69 datasets (251,083 studies), USA acquisition centers were involved. Chinese centers were involved in 9 datasets we surveyed (62,227 studies), then Netherlands and UK with 8 datasets and respectively 128,450 and 46,238 studies.

\begin{Insight}[boxed title style={colback=pink}]
Brain datasets involve additional requirements for the deployment of FMs such as appropriate licensing schemes and reliable collection and anonymization of the data. Only a subset of datasets matches the requirements for the design and deployment of large FMs.
\end{Insight}

\section{Best Foundation Model per Dataset}
\label{sec:best-archs}

It is quite hard today to identify the best architectures and models for each task and dataset, given the profusion of datasets and baselines evaluated in each paper, and the lack of standardized benchmarks and datasets in some settings. For example, among the 20 VQA foundation models, the most used benchmark for evaluation is \emph{Visual Question Answering in Radiology (VQA-RAD)} with 15 models evaluated on it, while among the 46 Segmentation foundation models, the most used benchmark is \emph{Brain Tumor Segmentation (BraTS)} with only 16 models using this dataset across its multiple releases (BraTS2018, BraTS2019, BraTS2021, etc.). There is no dataset used by more than five models for the other tasks covered in our study.

To identify the best models, we propose two methodologies. On the one hand, we analyze through a tournament which models beat which in their own evaluation (whatever the dataset used).
On the other hand, we focus on the most common benchmark and report the performance of the models from their own studies. In both cases, we are relying on the empirical evaluation in the respective works.

Then, we will introduce in more detail each of the models. Some models are highlighted multiple times, for different protocols or tasks, and are only described at their first occurrence.

\subsection{Tournament evaluation}

We differentiate in the following three types of models, (1) multimodal foundation models that support both text and images, and vision-only foundation models that can handle (2) 2D image data or (3) 3D volumetric image data. 

\begin{figure*}[!ht]
    \centering
    \vspace{-6.5em}
    \includegraphics[width=\linewidth]{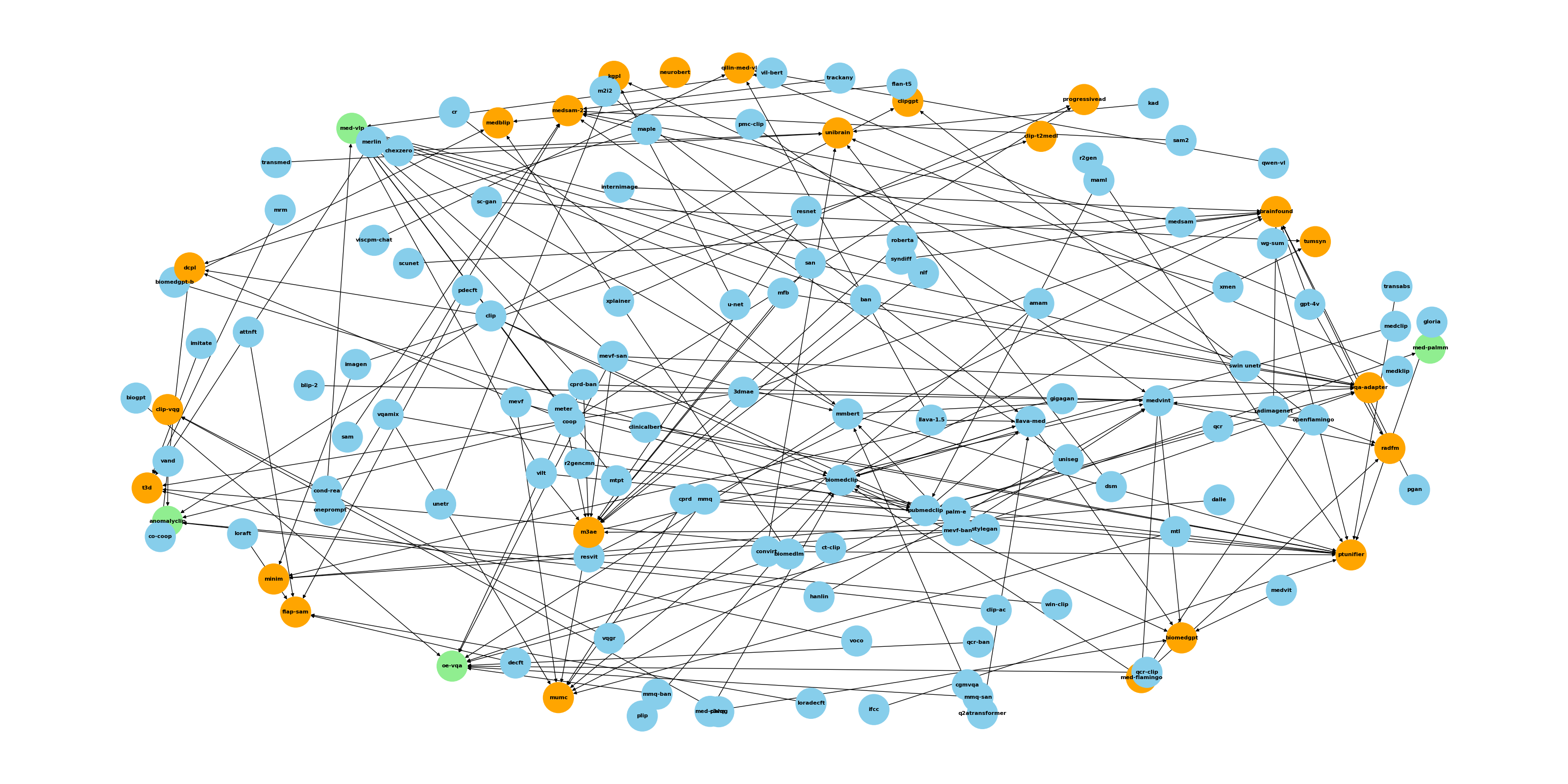}
    \vspace{-3.5em}
    \caption{Best Multimodal Foundation Models by tournament. In orange and green are the FM models covered, and in green are the best. A directed edge going from node A to node B means that model B is outperforming model A in one publication.}
    \label{fig:best_mfm}
\end{figure*}

\begin{figure*}[!ht]
    \centering
    \includegraphics[width=\linewidth]{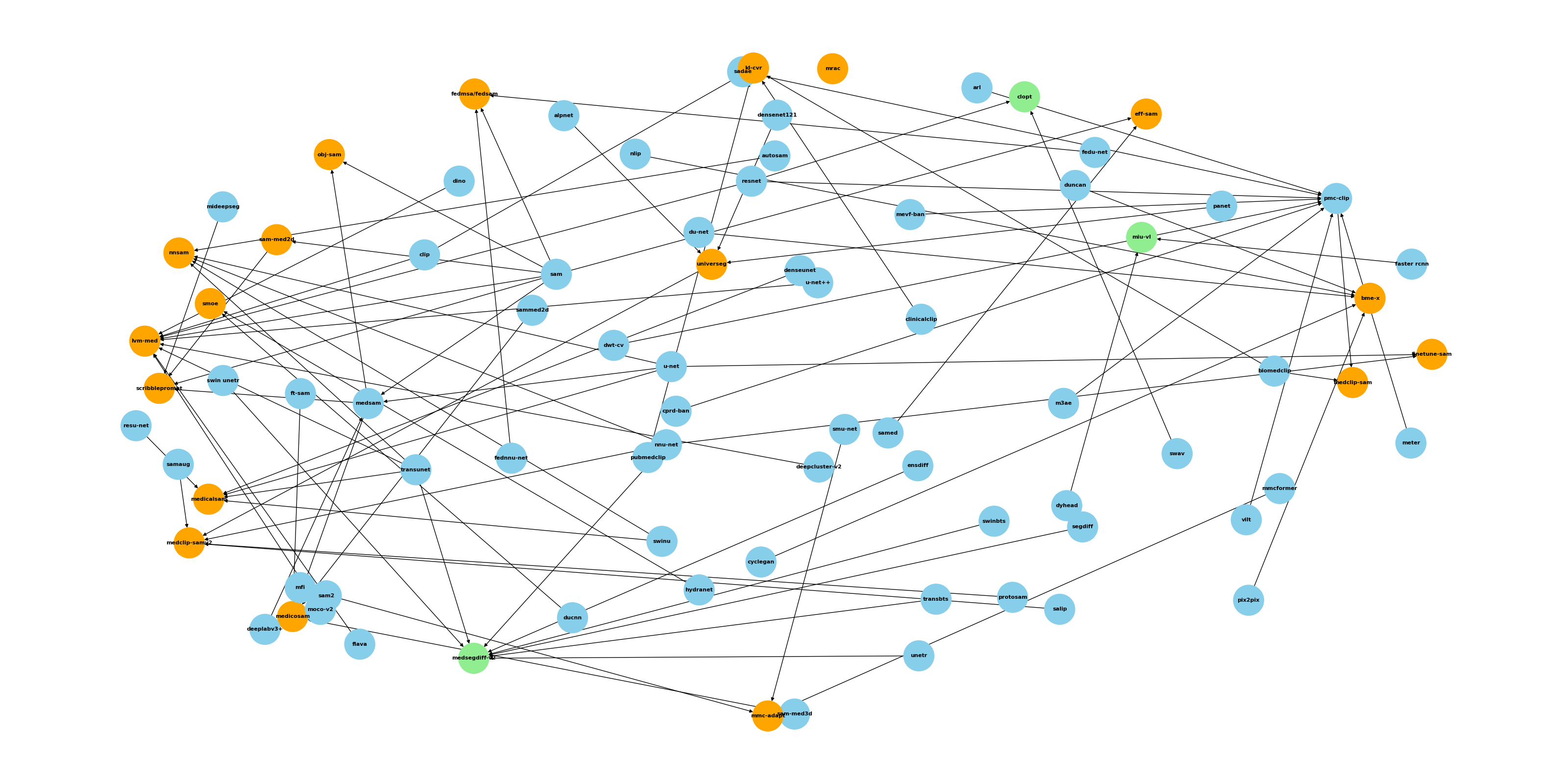}
    \vspace{-3.5em}
    \caption{Best 2D vision-only foundation models by tournament. In orange and green are the FM models covered, and in green are the best. A directed edge going from node A to node B means that model B is outperforming model A in one publication.}
    \label{fig:best_2dfm}
\end{figure*}

\begin{figure*}[!ht]
    \centering
    \vspace{-6.5em}
    \includegraphics[width=\linewidth]{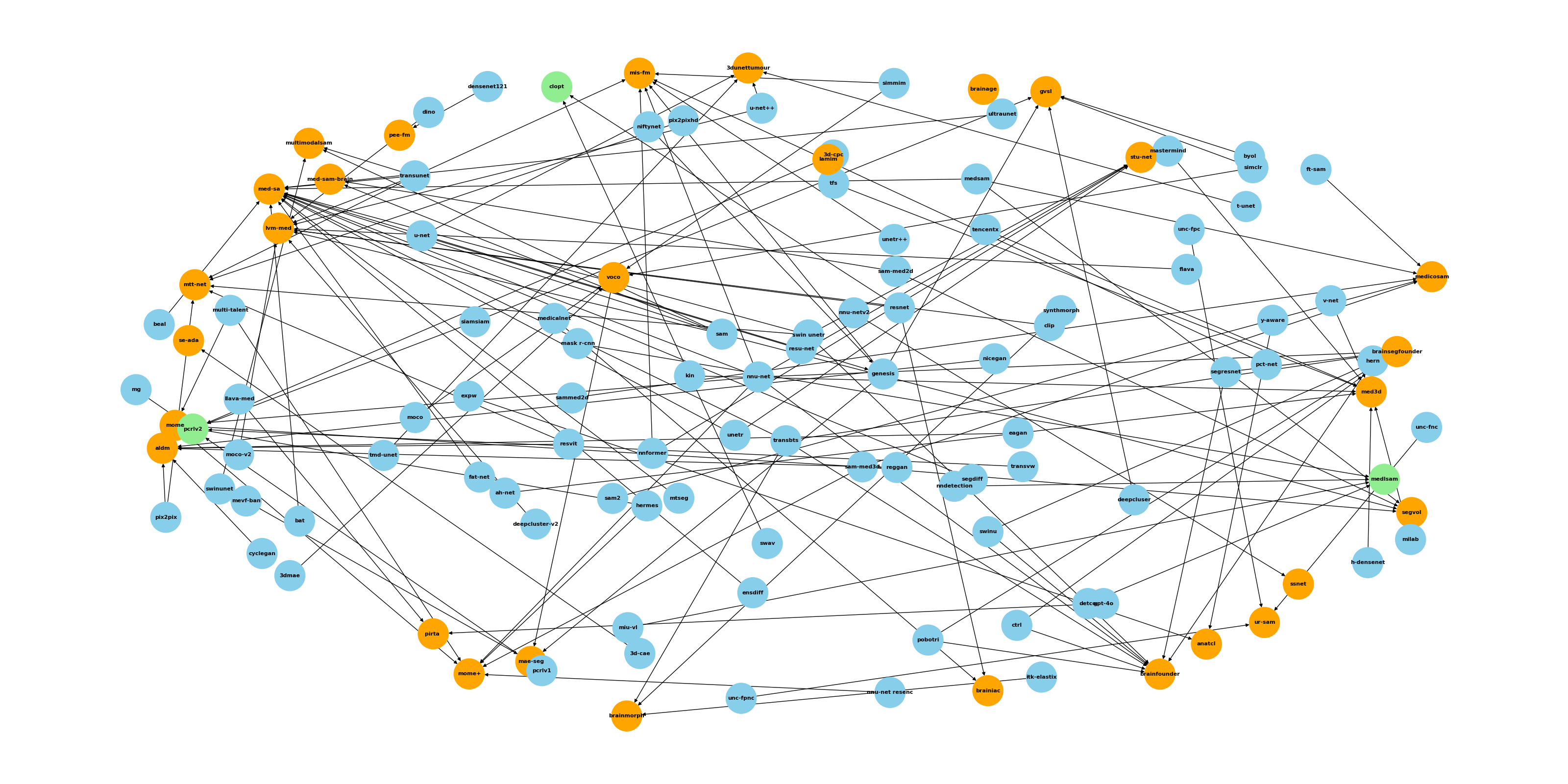}
    \vspace{-3.5em}
    \caption{Best 3D vision-only foundation models by tournament. In orange and green are the FM models covered, and in green are the best. A directed edge going from node A to node B means that model B is outperforming model A in one publication.}
    \label{fig:best_3dfm}
\end{figure*}

In each case, we build for multimodal FM in Fig.\ref{fig:best_mfm} (respectively Fig.\ref{fig:best_2dfm} for 2D vision FM, and Fig. \ref{fig:best_3dfm} for 3D vision FM) a graph of performance where we display in green and orange the foundation models considered in this section, and in blue the baselines that were beaten. An directed edge going from node A to node B means that model B is outperforming model A in one publication at least. In particular, we highlight in green the models that are peer-reviewed with more than 50 citations and that were not yet outperformed by any approach we surveyed. 

We highlight among multimodal FM four models: \textbf{OE-VQA} (58 citations)\cite{greenspan_oe-vqa_2023}, \textbf{Med-VLP} (65 citations)\cite{chen_med-vlp_2022}, \textbf{Anomaly-CLIP} (111 citations)\cite{zhou_anomalyclip_2024}, and \textbf{Med-PALMM} (353 citations)\cite{tu_med-palmm_2023}.
Three models match our criteria for vision-FM supporting 2D inputs: \textbf{MIU-VL} (82)\cite{qin_miu-vl_2023}, \textbf{CLOPT} (85)\cite{florin_c_ghesu_clopt_2022}, and \textbf{MedSegDiff-V2} (165)\cite{wu_medsegdiff-v2_2023}. Finally, three models match our criteria for vision-FM supporting 3D inputs: \textbf{PCRL-V2} (57)\cite{zhou_pcrlv2_2023}, \textbf{MedLSAM} (71)\cite{lei_medlsam_2024}, and \textbf{CLOPT} (85)\cite{florin_c_ghesu_clopt_2022}.

\subsection{Major benchmarks per task}

\paragraph{Semantic Segmentation: The BraTS benchmark}
BraTS is used in the evaluation of 24 FMs. While \citet{wu_radfm_2023}, \citet{tak_divyanshu_brainiac_2024} and \citet{lei_unibrain_2023} use it for classification, \citet{kim_jonghun_aldm_2024}, \citet{alan_q_wang_brainmorph_2024}, \citet{sun_yue_bme-x_2024} and \citet{han_xu_clip-t2medi_2025} use it for generative tasks. In Table \ref{tab:brats} we report the FM that provides a distinct evaluation score for BraTS glioma variants. We report the dice score claimed in the publications, converted from AUC if the original only uses AUC metrics. We also report which variants of the BRATS dataset were used in the evaluation. The results show that the best performing models are MoME and BrainSegFounder. It should be noted that MoME+ and MMC-adapt report performance with missing contrasts, Eff-SAM reports performance given small subsets of training data, and MAE-Seg evaluates Glioma segmentation performance with population shift. Another caveat of the comparison is that newer versions of BraTS contain more training data, which may provide unfair advantage when used. 

\begin{table}[ht]
\centering
\begin{tabular}{l|l|c}

\textbf{Model} & \textbf{Dataset version} & \textbf{DICE} \\
\hline
Med3D & 2013 & 55.41 \\
Med-SA & 2021 & 90.50 \\
Genesis & 2019 & 90.60 \\
MedSegDiff-V2 & 2021 & 90.80 \\
PCRLv2 & 2018 & 85.60 \\
LVM-Med & 2021 & 90.40 \\
Voco & 2019 & 78.53 \\
Eff-SAM & 2018 & 83.40 \\
BrainFounder & 2023 & 91.15 \\
MAE-Seg & Africa24 & 92.90 \\
MoME & 2021 & 92.10 \\
Med-SAM-Brain & 2023 & 61.85 \\
MMC-Adapt & 2021 & 86.90 \\
BrainSegFounder & 2023 & 91.15 \\
OBJ-SAM & 2023 & 91.90 \\
MoME+ & 2023 & 83.34 \\

\end{tabular}
\caption{FM evaluated on BraTS, with their dataset variants and reported DICE scores}
\label{tab:brats}
\end{table}

\begin{table}[!ht]
\centering
\small
\begin{tabular}{l|c|c|c|l}
\textbf{Model} & \textbf{Open} & \textbf{Close} & \textbf{Overall} & \textbf{Metric} \\
\hline
LLaVA-Med & 61.52* & 84.19 & -- & ACC \\
Med-PaLMM & -- & -- & 62.06 & F1 \\
Med-flamingo & -- & -- & 0.650 & BERT \\
BioMedCLIP & 67 & 76.5 & 72.7 & ACC \\
RadFM & -- & -- & 78.09 & F1 \\
PMC-Clip & 67 & 84 & 77.6 & ACC \\
MMBERT & 63.1 & 77.9 & 72 & ACC \\
M3AE & 67.23 & 83.46 & 77.01 & ACC \\
BiomedGPT & -- & -- & 73.2 & ACC \\
PubMedCLIP & 60.1 & 80 & 72.1 & ACC \\
Med-VLP & 67.7 & 86.76 & 79.16 & ACC \\
MUMC & 71.5 & 84.2 & 79.2 & ACC \\
PTUnifier & 68.7 & 84.6 & 78.3 & ACC \\
VQA-adapter & 66.1 & 82.3 & 75.8 & ACC \\
\end{tabular}
\caption{FM performances on VQA-RAD for close-ended and open-ended questions. *LLaVA-Med used Recall on open-ended questions.}
\label{tab:vqa-rad}
\end{table}

\paragraph{Question-answering: The VQA-RAD benchmark}

Although CLIP-VQG provides a combined performance over multiple VQA datasets, including VQA-RAD, 14 FMs provide a dedicated evaluation on this benchmark. We report these models in Table \ref{tab:vqa-rad}. Even within the same benchmark, we find inconsistencies in the metrics used, with two models reporting the F1 score (with RadFM achieving the best score of 78.09), and one model reporting a BERT similarity metric and blue score. 
The FMs with reported accuracy show close performances, with Med-VLP achieving the best accuracy in close-ended questions, and MUMC achieving the best performance in open-ended questions. We include the best-performing models in our following discussion.

\subsection{Discussion of the SoTA models}

Now that we identified the best FMs through two perspectives, we provide a summary of these models and highlight their major contributions.

\textbf{OE-VQA} introduces a transformer-based encoder-decoder architecture specifically designed for medical VQA. Unlike previous classification-based approaches, OE-VQA generates answers autoregressively rather than selecting from predetermined options.

\textbf{MED-VLP} explores how to inject expert knowledge to improve pre-training. First, it treats knowledge as an intermediate medium between vision and language modalities to which both are aligned, next, it injects knowledge into the multimodal fusion model to enable reasoning capabilities. Finally, it proposes knowledge-induced pretext tasks that guide the model training.

\textbf{Anomaly-CLIP.} Previous CLIP-based methods depend on object-specific or hand-crafted prompts and require labeled data or object semantics, limiting generalization. AnomalyCLIP’s object-agnostic approach removes these dependencies, allowing robust zero-shot performance and fine-grained anomaly localization across unseen domains.

\textbf{Med-PALMM.} The main contribution of this approach is the integration of clinical notes, imaging, and genomic data in one model. It leverages large-scale pre-training on diverse medical datasets to enable multi-task learning (segmentation, question-answering, etc).

\textbf{MIU-VL} introduces two automated prompt-generation methods that integrate medical knowledge and image-specific attributes, eliminating manual generic prompts. The prompts are augmented with domain-specific attributes (e.g., lesion texture) and expert terminology via pre-trained VQA model, medical language model, or a hybrid of both.

\textbf{CLOPT} demonstrates the importance of training at scale. It introduces a self-supervised learning framework combining contrastive learning, multi-modal feature clustering, and extensive data augmentations. It is trained on over 105 million medical 2D and 3D images across modalities (CT, MRI, X-ray, ultrasound) to address annotation scarcity and improve fine-tuning convergence speed.

\textbf{PCRLv2.} Unlike earlier SSL approaches that focus mainly on global invariance, PCRLv2 integrates pixel-level restoration and scale-aware learning to retain fine-grained and multiscale information. The framework’s non-skip U-Net and sub-crop strategies for 3D imaging further distinguish it by enhancing locality and scalability beyond what previous SSL models achieved.

\textbf{MedLSAM} introduces an automated 3D medical image segmentation pipeline by integrating MedLAM, a self-supervised localization model of a few shot, with the Segment Anything Model (SAM). Prior SAM-based medical adaptations required extensive slice-by-slice or point annotations, scaling annotation workload with dataset size, while MedLSAM achieves constant annotation effort.

\textbf{MoME} introduces a universal foundation model with specialized experts for different MRI modalities and a hierarchical gating network to combine predictions. It also introduces a curriculum learning strategy during the training phase to prevent the degeneration of each expert network and maintain their specialization.

\textbf{BrainSegFound} proposes a two-stage self-supervised pretraining strategy: first encoding healthy brain anatomy and then learning disease-specific features (e.g. tumors). Unlike previous supervised methods that need extensive labeled data, BrainSegFounder shows that one can leverage large-scale unlabeled datasets and then adapt the model to multiple modalities and downstream tasks.

\textbf{RadFM} introduces the first large-scale multimodal dataset with medical scans in 2D and 3D. The model adopts a generative approach rather than traditional discriminative methods. RadFM also introduces RadBench, a comprehensive evaluation benchmark comprising five tasks including modality recognition, disease diagnosis, and report generation for future research.

\textbf{MUMC.} Unlike earlier medical VQA models that used single objectives (e.g. contrastive loss or MLM), this work integrates multiple pre-training objectives to enhance cross-modal alignment and feature robustness. In addition, it extends the self-supervised pretraining with unimodal and multimodal contrastive losses.

\section{Discussion}
\label{sec-discussion}

\subsection{Major directions}

In the following, we highlight the major strategies adopted by the authors to improve their models. We identified four main areas in brain foundation model research. The first area of improvement involves efficiency. Some models aimed at designing new architectures that are cheaper at fine-tuning using adapters such as Low-Rank Adaptation (LoRA), and others designed lightweight architecture to reduce the inference cost using mixture of experts and shallow models. 

Other models investigated how to better adapt 2D models to a 3D setting. Instead of building from scratch foundation models for volumetric brains, they explored techniques to leverage pre-trained 2D models and adapt them to 3D data.

Some models went further and explored the adaption to a 4D setting, i.e., to fuse the learning of multiple imaging modalities or contrasts in one input, for example, leveraging T1W, T2W, Flair, and T1ce together to predict tumor segmentation. 

Another research direction also explored using non-image modalities, for instance, with \emph{smarter} prompts, or additional meta-data.

In parallel to extending inputs, other research explored ways to extend outputs to better support downstream tasks or improve generalization.  

\paragraph{Efficient fine-tuning}

A major strategy to extend a large textual model to medical imaging is the use of \emph{projection layers} and biomedical concept alignment. The approach converts image-text pairs into instruction-following data, for instance, image description. A linear projection layer is used to connect the vision encoder and the language model. During training, both the visual encoder and the LLM weights are frozen and only the projection layer is tuned. This approach is used as a first step for training LLAVA-MED\cite{li_llava-med_2023} for instance.

Another strategy is to fine-tune lightweight adapters interleaved with preexisting frozen layers. For example, MED-SA proposes a bottleneck adapter, consisting of a sequence of down-projection, ReLU activation, and up-projection. Each projection consists of a multilayer preceptron (MLP). Adapter strategies can be deployed to fine-tune both encoders and decoders \cite{wu_med-sa_2023,w_feng_eff-sam_2023,xiaoyu_shi_medicalsam_2024}, encoder alone \cite{cheng_sam-med2d_2023,greenspan_oe-vqa_2023,liu_qilin-med-vl_2023,chen_medblip_2023,gu_finetune-sam_2024,xiaoyu_shi_multimodalsam_2023,wang_sam-med3d_2024,liu_fedfms_2024,liu_progressivead_2024,xiaoliang_lei_mmc-adapt_2024,asokan_flap-sam_2024,archit_medicosam_2025}, or decoder alone \cite{lee_pirta_2024}.  

One subtype of adapters is the LoRA family. LoRA was first proposed by \cite{hu2021lora} as a parameter-efficient fine-tuning technique for large pre-trained models. Instead of updating all the parameters of a model during fine-tuning, LoRA introduces trainable low-rank matrices into certain layers (typically the attention and feed-forward layers) while keeping the original weights frozen. This approach significantly reduces the number of trainable parameters and the computational cost of adaptation. 

Let $W_0 \in \mathbb{R}^{d \times k}$ denote a pre-trained weight matrix in the model. In standard fine-tuning, the update to $W_0$ is a full-rank matrix $\Delta W$. LoRA constrains this update to be a low-rank matrix, parameterized as:
\begin{equation}
    \Delta W = BA,
\end{equation}
where $B \in \mathbb{R}^{d \times r}$, $A \in \mathbb{R}^{r \times k}$, and $r \ll \min(d, k)$ is the rank of the adaptation.
During the forward pass, the modified computation for an input $x$ is:
\begin{equation}
    h = W_0 x + \frac{\alpha}{r} BA x,
\end{equation}
where $\alpha$ is a scaling factor that controls the magnitude of the adaptation.

LoRA reduces the number of trainable parameters from $d \times k$ (full-rank update) to $r(d + k)$ (low-rank update). For example, for a layer with $d = k = 4096$ and $r = 8$, the number of trainable parameters drops from more than 16 million to just 65,536.

PIRTA used LoRA to fine-tune the LLaMA decoder, while MedicoSAM, FLAP-SAM, SAM-MED3D, Finetune-SAM, MedBLIP and OE-VQA used LoRA to fine-tune the encoder.

LoRA has inspired various extensions, such as dynamic rank allocation~\cite{lyn2024alora}, sparse low-rank adaptation~\cite{ding2023sora}, and federated heterogeneous adaptation~\cite{cho2024heterogeneous}. These methods further improve efficiency and adaptability in different scenarios. However, all the foundation models surveyed use the vanilla version of LoRA.

\paragraph{Efficient inference: MoE}

The mix of experts (MoE) was originally proposed in ~\cite{jacobs1991adaptive}, then popularized since ~\cite{lepikhin2020gshard}. It is a neural architecture that increases model capacity and efficiency by routing each input through a subset of specialized subnetworks, called experts. This conditional computation allows MoE models to scale to billions of parameters while keeping the computational cost per inference step manageable.

Each expert is typically a feed-forward neural network. The diversity among experts allows the model to specialize in different aspects of the input distribution. These experts are aggregated through a gating network that learns to route each input to the most appropriate experts. The top$k$ gating ensures sparsity and computational efficiency.

Let $x$ be an input and $n$ the number of expert networks $E_1, E_2, \dots, E_n$. The output of an MoE layer is a weighted sum of the experts' outputs, where the weights are determined by a gating network $G(x)$:
\begin{equation}
    h = \sum_{i=1}^n G_i(x) E_i(x),
\end{equation}
where $G(x) \in \mathbb{R}^n$ is a probability distribution over experts, typically computed as:
\begin{equation}
    G_i(x) = \frac{\exp(w_i^\top x)}{\sum_{j=1}^n \exp(w_j^\top x)},
\end{equation}
with $w_i$ being the trainable parameters of the gating network.

To increase efficiency, only the top-$k$ experts (usually $k=1$ or $2$) are activated for each input. The gating network selects these experts and assigns nonzero weights only to them. To prevent some experts from being over- or under-utilized, \cite{lepikhin2020gshard} introduced auxiliary losses to encourage balanced routing.

Notable extension to the MoE paradigm include 'hierarchical MoE', where multi-level expert routing is used for complex task decomposition, 'sparse MoE' that enforces fixed expert capacity buffers and are used in large language models such as Mixtral, and 'dynamic MoE' that leverages input-dependent expert count for multimodal or adaptive tasks.

Mixture of Experts with load balancing was implemented in 3 brain imaging foundation models; SMoE used MoE for lightweight encoders (the input was split into patches fed to different sparse experts), and MoME leveraged experts to handle different modalities (and fuse them). MoME+ extends the latter to handle missing modalities.

\paragraph{Adapting 2D pre-trained encoders to volumetric image encoders}

Among the 86 models surveyed in this study, 43 support volumetric inputs. Among them, 25 were trained from scratch using 3D encoders and 18 fine-tuned existing architectures to support volumetric input, that is, whole brain volumes instead of processing brains by slice. Three models (MedSAM-2, RadFM, and MedicoSAM) are dimension agnostic and support both 2D and 3D brain images as input. MedSAM-2 supports both dimensions by using SAM2'q video processing and considers volumetric images as video sequences on different axes. RadFM uses two encoders, one for 2D and one for 3D, which are converted into tokens, then interleaved with the textual tokens of the LLM. MedicoSAM is built on top of the encoders from SAMed for 2D inputs and MA-SAM for 3D inputs; the prompt encoder and the mask decoders are then fine-tuned to handle each of the cases. 



\paragraph{Supporting 4D inputs with incomplete sequences}

A main challenge in medical image datasets is the incomplete nature of the images recorded per patient and sequence. In fact, when a study aims to record multiple contrast protocols together (T1w, T1Gd, FLAIR, PET, etc.), some contrast must be discarded after quality checks. Mixing datasets of different nature leads to combining datasets that have different available contrasts. 
Traditionnaly, medical imaging models were trained on the smallest common set of available contrasts and protocols, but recent modality fusion approach allows more flexibility in the combinations.

RadFM for instance, by injecting the image modality as tokens interleaved with text tokens, can support a varied number of contrasts as input. MultiModamSAM introduces a cross-modality attention adapter to combine information from heterogeneous modalities. Unibrain supports heterogeneous contrasts by aligning the modality-wise imaging-report features, then projecting the concatenated features to the vision-language semantic space. BrainSegFounder pre-trained  the model using both T1 and T2 contrasts as different channels, but allows fine-tuning and inference on either T1 or T2 data alone by setting the two input channels to process the same type of data. LaMIM and PEE-FM use a more traditional masking strategy. By randomly masking random patches across all contrasts during pre-training, the model could adapt to completely missing contrasts at fine-tuning or inference. 

BrainFound and TumSyn support the use of text prompts to "describe" the imaging modality or contrast for each input. In this way, TumSyn was trained by mixing different conbinations of MRI contrasts, and BrainFound was trained by combining MRI contrasts and CT information.  MMC-Adapt specifically focuses on the missing modality scenario and proposes a twin network structure that processes missing and intact magnetic resonance imaging (MRI) modalities separately using shared parameters. While their image encoders are frozen, only multimodal adapters and spatial-depth adapters are fine-tuned. MoME+ introduces a dispatch network to support missing modalities. The dispatch network computes a soft dipatch that determines the weights to combine the available and missing inputs and generate the full multichannel input.

\paragraph{Augmenting the prompts}

14 foundation models introduce some automated way to generate or enrich the prompts fed to the encoders, Med-SA introduces a hyper-prompting adapter to enrich the representation of the prompt when combining the image and prompt embeddings. AnomalyCLIP proposes to learn object-agnostic text prompts that capture generic normality and abnormality in an image, thus improving the generalization and robustness to unseen prompts (in zero-shot anomaly detection scenarios). MIU-VL focuses on the design and automatic generation of medical prompts that can include expert-level knowledge and image-specific information. It combines a pre-trained VQA model querying on common properties (shape, color, location, ...) and Medical Language Model to infer these properties from the original prompt.

OA-VQA introduces a prompt-tuning mechanism that prepends a set of learnable tokens to the input prompt sequence, while MedSAM-2 leverages a memory bank mechanism to extend one prompt mask from one slice to propagate to all slices of the volume, achieving one-prompt segmentation capability from one slice.
ScribblePrompt introduces a prompt simulator in training interactive segmentation foundation models. The simulator allows the model to learn richer prompt combinations, such as line and centerline scribbles and contour annotations. UR-SAM also introduces a prompt augmentation mechanism that perturbs initial prompts, then uses uncertainty estimation to identify possible false-positive and false-negative regions.

DCPL improves prompt learning by generating domain-adaptive prompts for both visual and language branches. Each of the prompts is augmented by domain biases learned by training a lightweight neural network on top of the foundation model. SSNet proposes using specialized lightweight prompt-free models to generate coarse prompts for foundation models, and OBJ-SAM follows a similar intuition by using general-purpose object detection models to infer coarse prompts.  



\subsection{Pitfalls and blind spots}

\paragraph{Biases}
In medical imaging machine learning, bias refers to systematic errors that create a consistent gap between model predictions and the true clinical reality, often resulting from factors such as unbalanced datasets, annotation practices, or model development choices; these biases can lead to disparities in diagnostic accuracy and patient outcomes, especially among different demographic or clinical subgroups \cite{dir2024bias, pmc2024bias}.

Numerous studies have shown that machine learning (ML) models in medical imaging can exhibit biases, often manifesting as disparities in diagnostic performance between demographic or clinical subgroups \cite{ghamizi2022evaluating}. For example, AI models have been shown to learn spurious correlations related to age, sex, and race, leading to biased outcomes even when these attributes are not explicitly provided as input \cite{banerjee2023shortcuts}.

Only six foundation models explicitly considered potential biases in building their datasets and models. NeuroBERT for example, sampled a balanced dataset representative of different age (over 18), social strata, and ethnicity (40\% of non-white). It also built age-conditional embeddings with single-layer NN to capture age-dependent abnormalities. PEE-FM explicitly sampled a balanced dataset between genders (51\% male) and reported the size of the statistical effect for each population parameter. Similarly, BrainFound used a balanced MRI dataset (52\% male). CLIP-T2Med went further and placed gender bias as the core element of its evaluation. They proposed a new bias estimation metric to account for gender distribution and demonstrated that their approach significantly reduces potential gender biases. 

BrainAge, by definition, focuses on assessing the performance of the model across varying age populations (18-96 years). Its test dataset was also relatively balanced in gender (59\% male) and ethnicity (60\% white). Although Med-SAM-Brain did not enforce balance in training or test dataset, it explicitly provided evaluation results on pediatric brains and sub-saharian populations, in addition to the general population used in the training.    

\paragraph{Pathologies coverage}

The pathology distribution in foundation models is very different from the pathology distribution we identified by surveying the public datasets. Cancer is the most predominant family of pathologies covered in the foundation models with 34 models, vascular pathologies follow with 10 models, neurodegeneration with 10 models, neurodevelopmental disorders with five models, and mental health conditions with three models. Meanwhile, our analysis of the datasets in Section \ref{sec-datasets} showed that there are many more public studies on neurodegenerative diseases such as Parkinson and Alzeihmer than studies on cancer categories. We hypothesize that this mismatch reflects the large public cohorts of Alzheimer and Parkinson, providing a large volume of data, and thus leading to easier diagnosis and already very good performance of speciliazed traditional models. In fact, benchmarks for the diagnosis of Parkinson's and Alzheimer's ML such as BrainGB \cite{cui2022braingb} show that existing models achieve very good performance on these pathologies.

Rare brain cancers may require advanced models and remain an active topic of research in medical imaging, requiring the need for foundation models.    

The prevalence of pathologies introduced in Section \ref{sec:datasets} confirms the scarcity of brain cancers compared to vascular and neurodegenerative diseases. 

In general, our results suggest that FM is prioritizing only a subset of available medical imaging datasets, biased toward brain cancer research, and less reflective of the actual prevalence of brain pathologies, or available brain imaging datasets.

\paragraph{Performance evaluation}

One of the main insights of our study is the lack of commonly used benchmarks to evaluate brain FM, both in terms of tasks, datasets, and metrics. This gap makes it extremely difficult to compare new approaches and evaluate the generalization of different mechanisms between communities. In fact, new mechanisms are introduced to improve VQA that could lead to improved image synthesis or segmentation performance of models, and vice versa. The brain FM research is still scattered by task, use cases, and domain and is rarely studied in a holistic way.

Another challenge is the relevance of the metrics used. Although most of the models use traditional ML metrics such as accuracy, Area-under-Curve (AUC), Mean Square Error (MSE), Intersection over Union (IoU), Dice scores, or structural similarity index measure (SSIM). Only a handful investigated more medical metrics.

BME-X \cite{sun_yue_bme-x_2024} evaluated its performance also on raw metrics such as multiscale SSIM (MS-SSIM), universal quality index (UQI) and visual information fidelity (VIF). MS-SSIM extends SSIM by computing it across multiple scales. UQI models any image distortion as a combination of three factors: loss of correlation, luminance distortion, and contrast distortion. VIF evaluates the information shared between the reference and degraded images by exploring the connections between the information of the image and the visual quality. This metric demonstrates a higher correlation with radiologists' assessment of MR image quality compared to other metrics \cite{mason2019comparison}. In addition, to better evaluate the quality enhancement performance of their models, \citet{sun_yue_bme-x_2024} used the tissue contrast t-score (TCT)\cite{duffy2018retrospective}, which is defined as 

\begin{equation}
\mathrm{TCT} = \frac{|\mu_{\mathrm{wm}} - \mu_{\mathrm{gm}}|}{\sqrt{\sigma_{\mathrm{wm}}^2 + \sigma_{\mathrm{gm}}^2}},
\end{equation}
where $(\mu_{\mathrm{wm}}, \sigma_{\mathrm{wm}}^2)$ and $(\mu_{\mathrm{gm}}, \sigma_{\mathrm{gm}}^2)$ are the means and variances of white matter (WM) and grey matter (GM) intensities, respectively. 

WM and GM can be extracted and delineated with third-party tools such as the Infant Brain Extraction and Analysis Toolbox.
TCT measures the contrast between WM and GM, as well as the intensity variation within each tissue. The underlying hypothesis is that improved image quality should increase the contrast between GM and WM.

Similarly, FM can be evaluated with direct biological measurements, such as the estimation of error in the predicted volumes of brain structures or tumors.

The ultimate evaluation remains the human evaluation, with a Clinical Utility Score (CUS) or a Reader Study Agreement (RSA). CUS directly measures the real-world clinical impact and utility of the models' predictions to the relevant medical practitioners, while RSA reflects the consensus among multiple radiologists.

Among the 86 FMs reviewed, only 7 models involved a human evaluation:
Three experienced doctors from three different hospitals were involved in the evaluation of Brainfound.
Two human experts participated in the evaluation of MedSAM, Med-PaLMM reported an evaluation performed by four qualified thoracic radiologists. RadFM involved three radiologists with at least one year of clinical experience. They were asked to rate the quality of the model responses with scores from 0 to 5. The BiomedGPT generated responses were presented to a seasoned radiologist for scoring, while ScribblePrompt involved 16 neuroimaging researchers in an academic hospital for their human evaluation. 

MINIM involved five ophthalmologists who had at least 10 years of practice in retinal subspecialties and independently ranked each image and the corresponding report. Then, two independent senior retinal specialists 
verified the true labels and corresponding reports for each image.

\section{Limitations}
\label{sec-limitations}

The purpose of our review was to provide a thorough examination of the state of the literature on foundation models relevant to brain imaging. While our work is extensive, it faces multiple limitations that we attempted to curb and mitigate:

\paragraph{Rapidly Evolving Field}
The landscape of medical imaging foundation models is advancing at an exceptional pace. Our systematic review represents only a snapshot of the current state of research. Despite adhering to a rigorous research protocol, it is possible that some relevant developments or publications were missed. To address this limitation, we are committed to regularly updating the pre-print version of this review to incorporate the latest advancements and maintain its relevance.

\paragraph{Screening Process and PRISMA Compliance}
While we followed the PRISMA methodology to ensure transparency and reproducibility, the recommended practice of having two independent reviewers to evaluate and filter articles was not fully implemented; only one reviewer performed the initial screening and evaluation. However, all authors contributed to the subsequent discussion and analysis of the included studies, which helped mitigate potential biases introduced during the selection process.

\paragraph{Audience Accessibility and Scope}
We intentionally designed this review to be accessible and relevant to both the machine learning and medical imaging communities. This required providing detailed explanations of concepts that may be familiar to one community but not the other. Although this approach may introduce redundancy for some readers, it was a deliberate choice to foster cross-disciplinary understanding and engagement.

\paragraph{Caution in Meta-Analysis of Foundation Models}
The meta-analysis of leading foundation models presented in this review should be interpreted with caution. Our synthesis relies entirely on the figures and claims reported in the reviewed literature, some of which may not have undergone peer review or may not adhere to the highest methodological standards. As such, the comparative assessment of model performance may be influenced by inconsistencies in reporting practices and evaluation protocols across studies.


\section*{Conclusion}
\label{sec-conclusion}
The rapid evolution of foundation models (FMs) in brain imaging has unlocked new possibilities for diagnosing and managing neurological pathologies, yet critical challenges persist. Our analysis reveals that while brain FM research has accelerated since 2023, with architectures like Med3D, MedSAM, and LLaVA-Med leading innovation, the field remains constrained by a reliance on a narrow set of backbones (e.g., U-Net, SAM/CLIP encoders) and imbalanced task prioritization. For instance, MRI-based cancer segmentation dominates the literature, while PET imaging and mental health applications remain underexplored.

Data heterogeneity and quality issues pose significant barriers. Approximately 6\% of 3D imaging studies derive from duplicate datasets, risking data leakage during model training. Furthermore, label category overlaps and stark disparities in data availability across pathologies complicate benchmarking and generalization. Rare MRI contrasts (e.g., DCE) and PET tracers are particularly scarce. In addition, there is a stark contrast between the pathologies most covered in FMs (brain cancer), and the most prevalent pathologies in the population, and in available datasets (vascular and neurodegenerative pathologies).

Finally, FMs for brain imaging exhibit three critical limitations. First, biases in dataset composition and model design—such as imbalances in age, gender, or ethnicity—are rarely addressed, with only six models (e.g., NeuroBERT, CLIP-T2Med) explicitly implementing mitigation strategies like balanced sampling or bias-aware evaluation. Second, pathology coverage is skewed: while public datasets emphasize neurodegenerative diseases (e.g., Alzheimer’s, Parkinson’s), FMs disproportionately target brain cancers, neglecting conditions like mental health disorders despite available data. Third, evaluation practices lack clinical relevance: standardized benchmarks are absent, and only seven models incorporated human expert assessments (e.g., Med-SAM-Brain, RadFM), relying instead on traditional metrics like Dice scores that may not correlate with diagnostic utility. These gaps hinder the translation of FMs into clinically actionable tools.


\bibliographystyle{plainnat}
\bibliography{bib/general,bib/papers,bib/surveys,bib/2d_datasets,bib/3d_datasets,bib/intro,bib/def}



\end{document}